\documentclass[apj]{emulateapj}
\usepackage{amsmath}
\usepackage{array,multirow,natbib,color,tablefootnote}
\usepackage[flushleft]{threeparttable}
\usepackage{float}
\usepackage{graphicx}
\usepackage{subfigure}
\usepackage{setspace}
\usepackage{epstopdf,textcomp}

\newcommand{\bjdtdb}{\ensuremath{\rm {BJD_{TDB}}}}

\newcommand{\ms}{m s$^{-1}$}

\bibpunct[ ]{(}{)}{;}{}{}{,}
\shorttitle{Secondary Eclipse of XO-3b}
\shortauthors{Wong et al.}

\begin{document}
\title{Constraints on the Atmospheric Circulation and Variability of the Eccentric Hot Jupiter XO-3b}

\author{Ian Wong,\altaffilmark{1}\altaffilmark{*} Heather A. Knutson,\altaffilmark{1} Nicolas B. Cowan,\altaffilmark{2} Nikole K. Lewis,\altaffilmark{3,4} Eric Agol,\altaffilmark{5} Adam Burrows,\altaffilmark{6} Drake Deming,\altaffilmark{7} Jonathan J. Fortney,\altaffilmark{8} Benjamin J. Fulton,\altaffilmark{9} Jonathan Langton,\altaffilmark{10} Gregory Laughlin,\altaffilmark{8} Adam P. Showman\altaffilmark{11} }

\altaffiltext{1}{Division of Geological and Planetary Sciences, California Institute of Technology, Pasadena, CA 91125, USA}
\altaffiltext{*}{iwong@caltech.edu}
\altaffiltext{2}{Center for Interdisciplinary Exploration and Astrophysics (CIERA), Department of Earth \& Planetary Sciences, Department of Physics \& Astronomy, Northwestern University, 2145 Sheridan Road, Evanston, IL 60208, USA}
\altaffiltext{3}{Department of Earth, Atmospheric, and Planetary Sciences, Massachusetts Institute of Technology, Cambridge, MA 02139, USA}
\altaffiltext{4}{Sagan Fellow}
\altaffiltext{5}{Department of Astronomy, University of Washington, Seattle, WA 98195, USA}
\altaffiltext{6}{Department of Astrophysical Sciences, Princeton University, Princeton, NJ 08544, USA}
\altaffiltext{7}{Department of Astronomy, University of Maryland, College Park, MD 20742, USA}
\altaffiltext{8}{Department of Astronomy and Astrophysics, University of California at Santa Cruz, Santa Cruz, CA 95604, USA}
\altaffiltext{9}{Institute for Astronomy, University of Hawaii, Honolulu, HI 96822,
USA}
\altaffiltext{10}{Department of Physics, Principia College, Elsah, IL 62028, USA}
\altaffiltext{11}{Lunar and Planetary Laboratory, University of Arizona, Tucson, AZ 85721, USA}

\begin{abstract}
We report secondary eclipse photometry  of the hot Jupiter XO-3b in the 4.5~$\mu$m band taken with the Infrared Array Camera (IRAC) on the \textit{Spitzer Space Telescope}. We measure individual eclipse depths and center of eclipse times for a total of twelve secondary eclipses. We fit these data simultaneously with two transits observed in the same band in order to obtain a global best-fit secondary eclipse depth of $0.1580\pm 0.0036\%$ and a center of eclipse phase of $0.67004\pm 0.00013 $.  We assess the relative magnitude of variations in the dayside brightness of the planet by measuring the size of the residuals during ingress and egress from fitting the combined eclipse light curve with a uniform disk model and place an upper limit of 0.05\%. The new secondary eclipse observations extend the total baseline from one and a half years to nearly three years, allowing us to place an upper limit on the periastron precession rate of $2.9\times 10^{-3}$~degrees/day -- the tightest constraint to date on the periastron precession rate of a hot Jupiter.  We use the new transit observations to calculate improved estimates for the system properties, including an updated orbital ephemeris.  We also use the large number of secondary eclipses to obtain the most stringent limits to date on the orbit-to-orbit variability of an eccentric hot Jupiter and demonstrate the consistency of multiple-epoch \textit{Spitzer} observations.
\end{abstract}

\section{Introduction}\label{sec:intro}
The study of exoplanets has matured greatly in the past decade.  The \textit{Kepler} survey has enriched the body of known exoplanets with thousands of newly discovered objects \citep{petigura}. Meanwhile, the improved capabilities of both space- and ground-based telescopes have enabled the collection of high-precision photometry for the brightest of these planets. The majority of the observations carried out to date have focused on the class of gas giant planets known as hot Jupiters, which typically have orbital periods of a few days and atmospheric temperatures ranging between 1000 and 3000~K. These planets are some of the most favorable  targets for detailed study, as they have relatively frequent and deep eclipses \citep[e.g., ][]{winn}. 

By observing the secondary eclipse, which occurs when the planet passes behind its host star, we can deduce the shape of the planet's infrared emission spectrum and thereby study the properties of its dayside atmosphere \citep[e.g.,][]{charbonneau2005,deming}. The strong thermal emission of hot Jupiters makes them ideal targets for this type of analysis, and measurements of the secondary eclipse in multiple wavelength bands are commonly used to construct low-resolution  emission spectra at infrared wavelengths \citep[e.g.,][]{charbonneau2008,knutson2008}. By comparing these observations with theoretical atmospheric models, we can constrain the planet's atmospheric pressure-temperature profile, chemistry, albedo, and global circulation patterns \citep[see][for a recent review]{madhusudhan}. Infrared secondary eclipses have been measured for more than fifty exoplanets to date, of which the majority were obtained using the \textit{Spitzer Space Telescope} \citep[e.g.,][]{knutson2010,hansen}. Although the telescope's cryogen was exhausted in 2009, \textit{Spitzer}'s Infrared Array Camera \citep[IRAC; ][]{fazio} continues to operate in the 3.6 and 4.5~$\mu$m bands.

Hot Jupiters on eccentric orbits are of particular interest for atmospheric studies. In contrast to hot Jupiters that lie on circular orbits and are therefore likely tidally-locked, these eccentric bodies experience diurnal forcing due to their nonsynchronous rotation as well as significant variations in the incident stellar flux throughout an orbit, leading to time-variable atmospheric forcing \citep[][]{langton,iro,cowanagol,visscher,kataria}. The response of the planet's atmosphere reflects a balance between the time-varying stellar irradiation, the gravity wave propagation timescale, and the radiative time scale \citep[e.g.,][]{perez-becker}. These mechanisms can lead to thermal gradients in the planet's atmosphere, as well as temporal variations in those gradients. Unlike the circular case, phase curve observations of eccentric exoplanets cannot uniquely distinguish between flux variations due to the planet's rotation and variations due to the changing irradiation experienced by the planet.  Observations of the secondary eclipse ingress and egress break this degeneracy by providing a near-instantaneous picture of the planet's dayside temperature distribution. Specifically, the apparent offset of the center of eclipse phase due to a zonally-advected hot spot \citep{williams,agol},  along with deviations in the eclipse light curve morphology from a uniform disk model during ingress and egress \citep{rauscher,dewit,majeau}, can be used to construct a two-dimensional map of the planet's dayside brightness distribution and constrain important properties of the atmosphere. The only successful secondary eclipse mapping observation to date is of the circular hot Jupiter HD 189733b and was carried out independently by \citet{dewit} and \citet{majeau} using the 8~$\mu$m band on \textit{Spitzer}.

Recent general circulation models for eccentric hot Jupiters have shown that while the dayside atmospheric brightness of these planets may exhibit some spatial and temporal variability\citep{langton}, the phase curves of these planets should be relatively constant from one orbit to the next \citep{showman,lewis2010,cowanagol,kataria,heng}. The presence or absence of long-term variability has important implications for atmospheric studies of these planets.  It is common practice to combine data at different wavelengths from separate epochs years apart and assume that these data offer a consistent picture of the dayside emission spectrum. By comparing many individual secondary eclipse observations over a long time baseline, we can directly constrain the magnitude of orbit-to-orbit variations in the planet's atmospheric circulation.

In this paper, we analyze twelve secondary eclipse and two transit observations of the eccentric hot Jupiter XO-3b obtained in \textit{Spitzer}'s 4.5~$\mu$m micron band. This planet has a mass of  $M_{p} = 11.67 \pm 0.46~M_{\mathrm{Jup}}$ \citep{johnskrull,hirano} and orbits its host star \citep[spectral type F5V, $R_{*}=1.377\pm0.083 R_{\Sun}$, $T_{*} = 6759 \pm 79$~K, and $\log{g} = 4.24\pm 0.03$;][]{winn,torres} with a period of 3.19~days and an orbital semi-major axis of $a=0.045$~AU \citep{winn}. With an  orbital eccentricity of $e = 0.2833\pm 0.0034$ \citep{knutson2014}, the stellar flux received by this planet varies by a factor of 3.2 between periapse and apoapse. \cite{machalek} reported secondary eclipse depths for this planet in the 3.6, 4.5, 5.8, and 8.0 micron Spitzer bands, but these data were taken in stellar mode and as a result have a lower cadence and precision than the now-standard staring mode observations. Here, we present twelve additional secondary eclipse and two transit observations with a total baseline of almost three years.

The paper is organized as follows: The observations and methodology for extracting photometry are described in Section~\ref{sec:obs}. In Section~\ref{sec:analysis}, we present the best-fit eclipse parameters for individual secondary eclipse and transit observations as well as the combined eclipse light curve. We then use our results to obtain an updated orbital ephemeris and discuss the implications of our results for the planet's atmospheric dynamics in Section~\ref{sec:dis}.

\section{Observations and photometry}\label{sec:obs}

A total of thirteen secondary eclipse  and two transit observations of XO-3b were carried out using the Infrared Array Camera (IRAC) on the \textit{Spitzer Space Telescope} in the 4.5~$\mu$m channel. For all observations, we utilized the IRAC subarray mode, which produced 32 $\times$ 32 pixel (39$''\times$39$''$) images with 2.0~s integration times. The observations include ten individual secondary eclipse observations, each of which contains 14912 images over 8.4~hr; additionally, two secondary eclipses and one transit are contained within a full-orbit observation, which was obtained on UT 2013 May 5-8 and has a total duration of 86.7~hr, corresponding to 153600 images (PID: 90032). The long duration of observation and limited on-board memory necessitated multiple breaks for downlinks. One of these downlinks occurred during the second of the two secondary eclipses in the full-orbit observation, which led to difficulties during our analysis. Therefore, this eclipse observation is omitted from the analysis presented here. For the other secondary eclipse and transit contained in this full-orbit observation, we extracted 14912 images from the full-orbit observation spanning the duration of each event to use in our analysis. This number was chosen to match the lengths of the other ten individual eclipse observations; we obtain consistent results when including more or fewer images. The remaining secondary eclipse and transit are contained within a 66~hr observation obtained on UT 2012 April 5-8 (PID: 60058); as with the full-orbit observation, we extracted 14912 images spanning the secondary eclipse and transit for use in our analysis. In Table~\ref{tab:basic}, we list the start times of the twelve secondary eclipse and two transit observations that are included in our analysis.

\begin{table*}[t!]
\caption{Eclipse/transit observation start times and optimal aperture photometry} \label{tab:basic}
\centering
\renewcommand{\arraystretch}{1.2}
\begin{tabular}{c c c |m{0.001cm} c m{0.001cm} c  }
\hline\hline

&$\#$    &       Observation Start Time &  &Median Photometric Aperture & & Type \\
 &&  (BJD$_{\mathrm{TT}}-$2455000) &  & (pixels) &  &  \\
\hline
 & & & & & &  \\
\parbox[t]{1mm}{\multirow{12}{*}{\rotatebox[origin=c]{90}{Eclipses}}}&1 &    294.36810 & & 2.40    & &fixed   \\
&2 & 1242.25466 & & 3.37 & &scale - 1.75  \\
&3 & 1248.66323 & & 2.78 & &shift - 1.2 \\
&4 & 1251.83395 & & 3.45 & &scale - 1.75\\
&5 & 1255.03213 & & 2.98  & &shift - 1.1  \\
&6 & 1264.60591 & & 2.71 & &scale  - 1.70 \\
&7 & 1270.99427 & &3.53  & &scale - 1.65 \\
&8 & 1405.03292 & & 2.41 & &scale - 1.55 \\
&9 & 1417.74069 & &3.28  & &scale - 1.80 \\
&10 & 1430.56660 & & 2.83 & &scale - 1.55\\
&11 & 1433.75742 & & 2.82 & &shift - 0.8  \\
&12 & 1436.94105 & & 3.26 & &scale - 1.95 \\
& & & & & &\\
\hline
\parbox[t]{1mm}{\multirow{4}{*}{\rotatebox[origin=c]{90}{Transits}}}& & & & & &\\
&1 &   292.22203  & &  2.20   & &fixed   \\
&2 & 1418.83774 & &  3.52 & &scale - 1.90\\
&& && &&\\
\hline\hline
\end{tabular}
\end{table*}

Data from previous observations using the IRAC instrument displayed a short-duration, ramp-like behavior at the start of each observation likely due to the settling of the telescope at a new pointing position \citep[e.g.,][]{knutson2012}. To address this, each individual eclipse observation was preceded by a $\sim$30 minute ``peak up'' observation, which allows the telescope pointing to stabilize before making a final position adjustment to place the star in the middle of the central pixel in the subarray.  We exclude these initial peak-up observations from our subsequent analysis. We find that our post-peak-up light curves do not display any obvious ramp-like behavior. Nevertheless, we experimented with trimming the first 10 or 30 minutes of data before fitting to our model eclipse light curve and obtained consistent results with no significant improvement in the residual scatter. In the analysis presented here, we utilize the full eclipse observations.

We extract photometry using methods similar to those used in previous secondary eclipse analyses \citep[e.g.,][]{todorov,orourke}. The basic calibrated data (BCD) files are dark-subtracted, flat-fielded, linearized, and flux-calibrated using version S19.1.0 of the IRAC pipeline. The exported data from a secondary eclipse or transit observation comprise a set of 233 FITS files, each with 64 images and a UTC-based Barycentric Julian Date (BJD$_{\mathrm {UTC}}$) time stamp designating the start of the first image. The mid-exposure time stamp for each individual image in a FITS file is calculated assuming uniform spacing and using the difference between the AINTBEG and ATIMEEND headers, which indicate the start and end of each 64-image series. We then transform each BJD$_{\mathrm {UTC}}$ time stamp into the Barycentric Julian Date based on the Terrestrial Time standard (BJD$_{\mathrm {TT}}$), using the conversion at the time of our observations \citep{eastman}. The continuous BJD$_{\mathrm {TT}}$ standard is preferred because leap seconds are occasionally added to the BJD$_{\mathrm {UTC}}$ standard. 

To estimate the sky background, we create a histogram of all pixel values in each image and fit a Gaussian function. We avoid contamination from the wings of the star's point-spread function (PSF) by excluding pixels within a radius of 15 pixels from the center of the image, as well as the 13th-16th rows and the 14th and 15th columns, where the stellar PSF extends close to the edge of the array. In addition, we exclude the top (32nd) row of pixels, since they have values that are consistently lower than those from the rest of the array. Before binning the remaining pixel values and fitting a Gaussian, we iteratively trim outlier values that are more than $3\sigma$ from the median value. After subtracting the best-fit sky background from the subarray images, we correct for transient ``hot pixels''  in each set of 64 images by comparing the intensity of each pixel to its median value. Pixel intensities varying by more than $3\sigma$ from the median value are replaced by the median value. The average percentage of pixels replaced across all pixels contained within an eclipse/transit observation is less than $0.35\%$.

The position of the star in each image is determined using flux-weighted centroiding \citep[see, for example,][]{knutson2008,charbonneau2008}: Defining a parameter $r_{0}$, we iteratively calculate the flux-weighted centroid within a circular region of radius $r_{0}$ pixels centered on the estimated position of the star. We optimize $r_0$ separately for each data set to minimize the scatter in the residuals from our eventual best-fit light curve (Section~\ref{sec:analysis}). For each image, we also estimate the width of the star's PSF by computing the noise pixel parameter \citep{mighell}, which is defined in Section~2.2.2 of the \textit{Spitzer/IRAC} instrument handbook as
\begin{equation}\widetilde{\beta}=\frac{\left(\sum_{i}I_{i}\right)^2}{\sum_{i}I_{i}^2},\end{equation}
where $I_{i}$ is the intensity detected in the $i$th pixel. We define the parameter $r_{1}$ as the radius of the circular aperture used to calculate $\widetilde{\beta}$ and optimize the value for each individual eclipse/transit to minimize the scatter in the residual of the best-fit light curve from each individual fit to an eclipse or transit (Section~\ref{subsec:fits}).

We perform aperture photometry to calculate the flux in each image, using either a fixed radius ranging from 2.0 to 5.0 in pixels, or a time-varying radius equal to the square-root of the noise pixel parameter $\widetilde{\beta}$, with either a constant scaling factor or a constant shift \citep[see][for a full discussion of the noise-pixel-based aperture]{lewis}. To determine the optimal choice of aperture photometry for each eclipse or transit, we subtract our best-fit model light curve from each photometric time series and compute the RMS scatter in the resultant residuals, binned in five-minute intervals, and pick the version of the photometry that gives the minimum scatter. The binning is chosen so as to assess scatter in the data over time scales comparable to the duration of ingress and egress (21.8 minutes), which is more likely to affect the results of the light curve fits. We found that for 11 of the 12 secondary eclipses and one of the two transits, choosing the time-varying aperture photometry yielded the smallest residual scatter. The medians of the time-varying radius for the photometric aperture and types of aperture (fixed, scale, or shift) used for all data sets are listed in Table~\ref{tab:basic}. For the case of time-varying aperture, the scaling factor or number of pixels shifted is also listed. 

It is important to note that by choosing the optimal aperture through minimizing the resulting RMS scatter, we may be partially fitting away any deviation during ingress and egress from our eclipse model due to non-uniform dayside brightness. In particular, the pixel-mapping method we use to estimate and remove the instrumental noise (Section~\ref{subsec:correction}) eliminates as much structure as possible from the raw light curve, with no distinction made between astrophysical signals and noise due to intrapixel sensitivity variations. However, there are many flux measurements at a given pixel location before and after ingress and egress, which statistically dominate the pixel map calculation. Moreover, we obtain consistent results using a polynomial noise decorrelation function in the $x$ and $y$ position of the star centroid, a method that is not capable of removing such small-scale structure in the light curve during ingress and egress. Ultimately, our lack of \textit{a priori} knowledge of the precise center of eclipse time is much more important in potentially fitting away any astrophysical signal during ingress and egress (see Section~\ref{subsec:circulation}).

Before fitting to the model, we iteratively filter out points in our light curves with uncorrected measured fluxes that vary by more than 3$\sigma$ from the median values in the adjacent 64 frames (i.e., the length of one FITS file) in the time series. Choosing a larger or smaller interval for computing the median values does not significantly affect the number of excised points. For each of the twelve eclipses, the percentage of data points removed is less than 0.5$\%$.

\section{Data analysis}\label{sec:analysis}
\subsection{Orbital parameters}\label{subsec:ephe}
The routines for fitting the secondary eclipse and transit light curves require precise values for the planet's orbital parameters. In particular, we need the planet's orbital period $P$, inclination $i$, eccentricity $e$, scaled semi-major axis $a/R_{*}$, and pericenter longitude $\omega$, as well as the ratio between the planetary and stellar radii $R_{p}/R_{*}$. When optimizing our choice of photometric aperture, we take values from \citet{winn} for all orbital parameters except eccentricity and pericenter longitude, for which we take values from \citet{knutson2014} ($e=0.2833$ and $\omega=346.8^{\circ}$). In the final version of our fits of individual eclipses and transits, we use the updated values for $i$, $R_{p}/R_{*}$, and $a/R_{*}$ in Table~\ref{tab:best} computed from our global fit (see Section~\ref{subsec:fits}).

\subsection{Secondary eclipse/transit model}\label{subsec:model}
We calculate the transit and eclipse light curves using the formalism of \citet{mandelagol} and include free parameters for the eclipse/transit depth $d$ and the center of eclipse/transit time $t_{0}$.  In addition, we approximate the planet's phase curve in the region of the secondary eclipse or transit as a linear function of time where the slope (``phase constant") is a free parameter in our fits. Because this phase curve slope is absent during the secondary eclipse, we keep the observed flux between second and third contacts constant at $(1-d)$ and scale the amplitudes of ingress and egress appropriately to match them to the out-of-eclipse phase curve. For fits to individual secondary eclipses we allow the depths and times to vary as free parameters along with the phase constant, while fixing $i$, $a/R_{*}$, and $R_{p}/R_{*}$. In the case of  transits, the depth $d$ is equal to the square of the radius of the planet relative to that of the host star, $R_{p}/R_{*}$, which we allow to vary as a free parameter while keeping $i$ and $a/R_{*}$ fixed.  In our global fits, which include both the secondary eclipse and the transit light curves, we allow all of these parameters to vary freely in the fit. The host star XO-3 has an effective temperature $T_{*} = 6759 \pm 79$~K and a specific gravity $\log{g} = 4.24\pm 0.03$ \citep{torres}. When fitting to the transit light curves, we use a four-parameter non-linear limb-darkening law \citep{sing} with parameter values calculated for a star with $T_{*}=6750$~K and $\log{g}=4.0$ ($c_{1}=0.4885$, $c_{2}=-0.7003$, $c_{3}=0.7724$, and $c_{4}=-0.3102$).

\subsection{Correction for intrapixel sensitivity variations}\label{subsec:correction}
The fluxes measured within our photometric aperture display a strong correlation with variations in the position of the star on the array. This effect is due to the non-uniform sensitivity of an individual pixel across its area \citep{charbonneau2005} and must be accounted for when fitting the fluxes with the eclipse/transit model. In our analysis, we correct for this systematic effect in two different ways.  

First, we adopt an approach similar to that used in previous studies \citep[e.g.,][]{todorov} and include a second order polynomial function of the $x$ and $y$ positions of the star as part of the eclipse fit:
\begin{equation}\label{pixel}I(\lbrace c_{i}\rbrace,x,y)=c_{0}+c_{1}x+c_{2}y+c_{3}x^2+c_{4}y^2+c_{5}xy.\end{equation}
We evaluate the utility of individual terms in this expression using the Bayesian information criterion (BIC):
\begin{equation}\label{BIC}{\mathrm {BIC}}=\chi^2+k\ln(n),\end{equation}
where $k$ is the number of free parameters in the model (including the parameters in the eclipse model), $n$ is the number of data points, and $\chi^2$ is the standard metric for goodness of fit. Using this metric, we find that the constant and linear terms in the decorrelation function suffice for eclipses 1, 3, 7, and 10. The BIC is minimized by including the $x^2$ term for eclipses 5 and 12, by including the $y^2$ term for eclipses 2, 5, 6, 8, and 9, and by including all the second-order terms for eclipse 11. It is important to note that the intrinsic error in the light curve, which formally enters into the calculation of $\chi^2$, is unknown, and our $\chi^2$ values are based solely on the spread in the residuals from the fits. However, the large number of data points available ensures we can obtain a reasonably accurate estimate of the noise in each data set, with a correspondingly small uncertainty in our calculation of the BIC. We obtain consistent results for the eclipse depths and times using any decorrelation polynomial of at least first order.

\begin{table*}[t!]
\caption{Parameters for Individual Transit and Secondary Eclipse Fits} \label{tab:values}
\centering
\renewcommand{\arraystretch}{1.2}
\begin{tabular}{c c c m{0.001cm} c m{0.001cm} c m{0.001cm} c  }
\hline\hline

&$\#$    &       Eclipse/Transit & & Eclipse/Transit&  &Eclipse/Transit & & Phase \\
& & Center Time & & Phase & & Depth & & Constant\\
& & $t_{0}$ & & $\phi_{0}$ & & $d$ & & $c$ \\
& &  (BJD$_{\mathrm{TT}}-$2455000) & & & & ($\%$) &  &($\times 10^{-4} $~d$^{-1}$)  \\
\hline
\\
\parbox[t]{1mm}{\multirow{12}{*}{\rotatebox[origin=c]{90}{Eclipses}}} &1 & 294.5736$^{+0.0014}_{-0.0016}$& & 0.67067$^{+0.00045}_{-0.00050}$& &0.1722$^{+0.0130}_{-0.0157}$ & & $-10.4^{+6.8}_{-7.5}$\\
&2 &  1242.4563$\pm 0.0010$ &  & 0.66987$^{+0.00033}_{-0.00031}$ & & 0.1732$^{+0.0085}_{-0.0084}$& & 3.0$\pm 5.5$\\
&3              & $1248.8405^{+0.0010}_{-0.0009}$ &  & 0.67022$^{+0.00030}_{-0.00029}$ & &0.1653$^{+0.0084}_{-0.0083}$ & & 2.5$\pm 3.6$ \\
&4              & $1252.0303^{+0.0014}_{-0.0011}$ &  & 0.66968$^{+0.00043}_{-0.00033}$ & &0.1561$^{+0.0085}_{-0.0084}$& & 17.2$^{+5.3}_{-4.8}$ \\
&5              & 1255.2215$\pm 0.0012$ &  & 0.66957$^{+0.00037}_{-0.00036}$ & &0.1467$\pm 0.0080$ & & $-$7.9$^{+5.0}_{-4.8}$ \\
&6              & $1264.7988^{+0.0013}_{-0.0016}$ &  & 0.67042$^{+0.00041}_{-0.00050}$ & & 0.1614$^{+0.0111}_{-0.0091}$& & $-$5.6$^{+6.3}_{-5.7}$ \\
&7             & $1271.1801^{+0.0010}_{-0.0009}$  &  & 0.66987$^{+0.00030}_{-0.00029}$ & & 0.1587$^{+0.0088}_{-0.0120}$& &  23.0$\pm 6.4$\\
&8              & 1405.2229$\pm 0.0011$ &  & 0.66937$\pm 0.00036$ & &0.1533$^{+0.0080}_{-0.0083}$ & &  13.4$\pm 5.4$\\
&9              & $1417.9918^{+0.0014}_{-0.0011}$ &  & 0.67024$^{+0.00043}_{-0.00035}$ & & 0.1585$\pm 0.0080$& & $-$2.9$\pm 4.1$ \\
&10              &  $1430.7555^{+0.0013}_{-0.0012}$&  & 0.66948$^{+0.00040}_{-0.00038}$ & & 0.1578$^{+0.0130}_{-0.0061}$& & 15.2$^{+6.2}_{-5.7}$ \\
&11              & 1433.9484$^{+0.0010}_{-0.0011}$ &  & 0.66991$^{+0.00030}_{-0.00033}$ & &0.1565$^{+0.0081}_{-0.0080}$ & & 17.0$\pm 4.3$  \\
&12              & $1437.1415^{+0.0012}_{-0.0013}$ &  &  0.67040$^{+0.00036}_{-0.00040}$ & &0.1468$^{+0.0080}_{-0.0082}$ & & 11.0$^{+4.1}_{-4.2}$  \\
\\
\hline

\parbox[t]{1mm}{\multirow{4}{*}{\rotatebox[origin=c]{90}{Transits}}}&& &&($\times 10^{-4}$) &&\\
&1 & 292.4320$\pm 0.0003$& & $-$2.08$\pm 0.81$& & 0.7830$^{+0.0085}_{-0.0088}$& &$-$5.3$^{+4.9}_{-5.8}$ \\
&2 &  1419.0441$\pm 0.0003$ &  & 1.10$^{+0.81}_{-0.5}$ & & 0.7760$^{+0.0086}_{-0.0088}$& & 7.8$^{+5.2}_{-5.3}$\\
\\
\hline\hline
\end{tabular}
\end{table*}

The second approach we use to remove the intrapixel sensitivity effect is to create a map of the pixel response. Our method is similar to the one described in \citet{ballard}. We approximate the star in an image $j$ as a point source at the location on the array given by the measured centroid position $(x_{j},y_{j})$ and model the sensitivity of that location by comparing other images with measured centroid positions near $(x_{j},y_{j})$. The effective pixel sensitivity at a given position is calculated using
\begin{equation}\label{pixelmap}F_{\mathrm{meas},j}=F_{0,j}\sum\limits_{i=0}^{m}e^{-(x_{i}-x_{j})^2/2\sigma_{x,j}^2}\times e^{-(y_{i}-y_{j})^2/2\sigma_{y,j}^2},\end{equation}
where $F_{\mathrm{meas},j}$ is the flux measured in the $j$th image, $F_{0,j}$ is the intrinsic flux, $x_{j}$ and $y_{j}$ are the measured $x$ and $y$ positions of the star centroid, and $\sigma_{x,j}$ and $\sigma_{y,j}$ are the standard deviations of the $x$ and $y$ over the full range in $i$, reflecting the relative spread in position spanned by the points included in the sum. In essence,  we are implementing an adaptive smoothing technique where the spatial resolution is smaller in more densely-sampled regions. The index $i$ in Eq.~\eqref{pixelmap} sums over the nearest $m=50$ neighbors with distance defined as $d_{i,j} = (x_{i}-x_{j})^2+(y_{i}-y_{j})^2$. We chose this number of neighbors to be large enough to adequately map the pixel response and also low enough to remain relatively computationally efficient \citep{lewis}.

The advantage of using the pixel-mapping approach is that it uses the measured fluxes themselves to account for the intrapixel sensitivity effect and does not require additional parameters to be fitted, as is the case with the polynomial approximation. This is especially helpful in our global fit to all twelve secondary eclipses and two transits, for which the use of decorrelation functions incurs a prohibitively large computational overhead. Comparing the fitted eclipse parameters for individual eclipses obtained using both approaches, we see that the values and relative uncertainties are consistent. We therefore use the pixel-mapping technique in the final version of our analysis.

\subsection{Parameter fits}\label{subsec:fits}

\begin{figure*}[t*]
\begin{center}
\includegraphics[width=19cm]{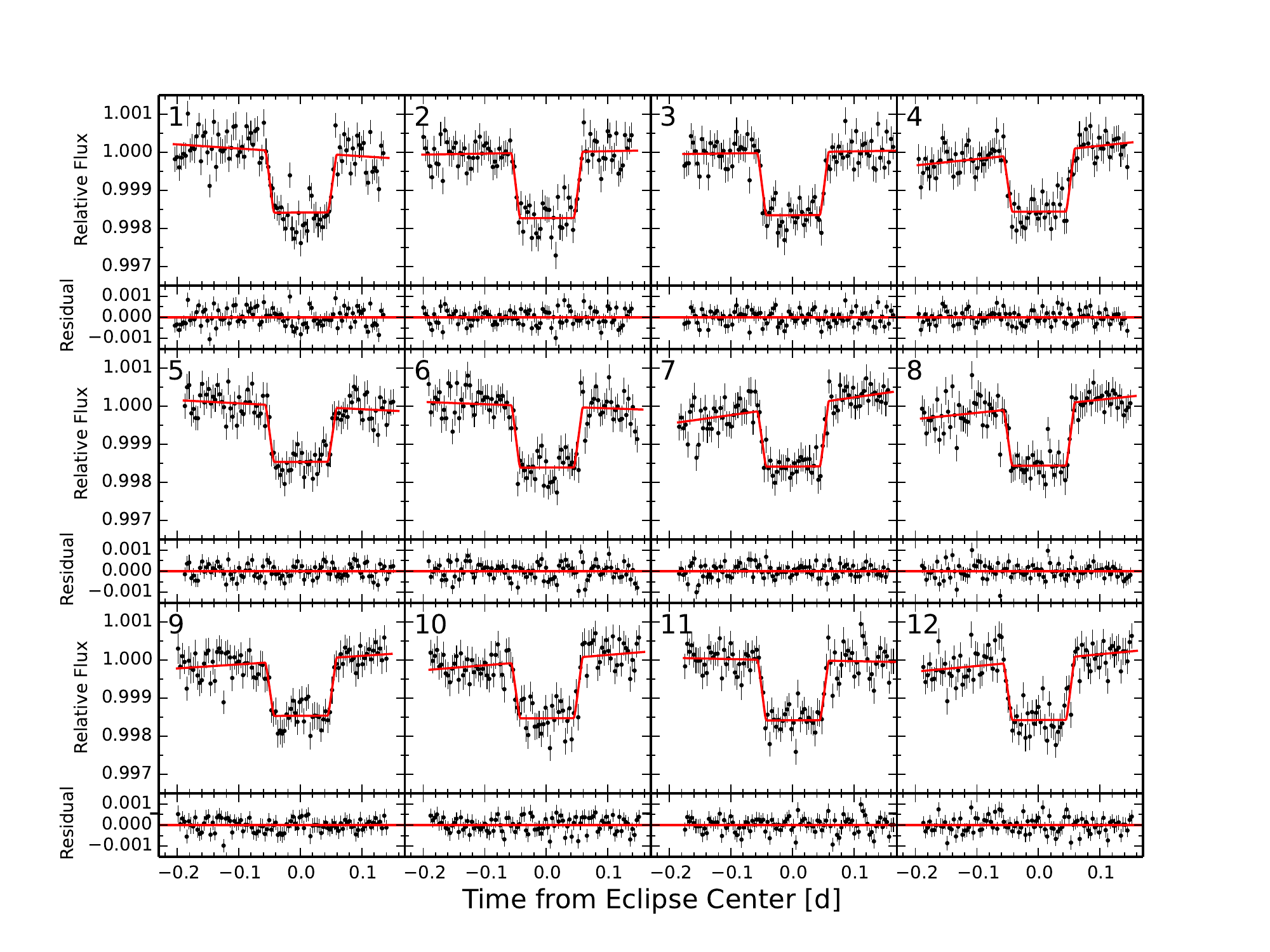}
\end{center}
\caption{Plot of twelve secondary eclipses observed in the 4.5~$\mu$m \textit{Spitzer} band, binned in three-minute intervals, with intrapixel sensitivity effects removed (black dots with error bars) and best-fit light curves overplotted (solid red lines). The residuals from each fit are shown below the corresponding light curve.} \label{eclipses}
\end{figure*}

\begin{figure}[t*]
\begin{center}
\includegraphics[width=9.4cm]{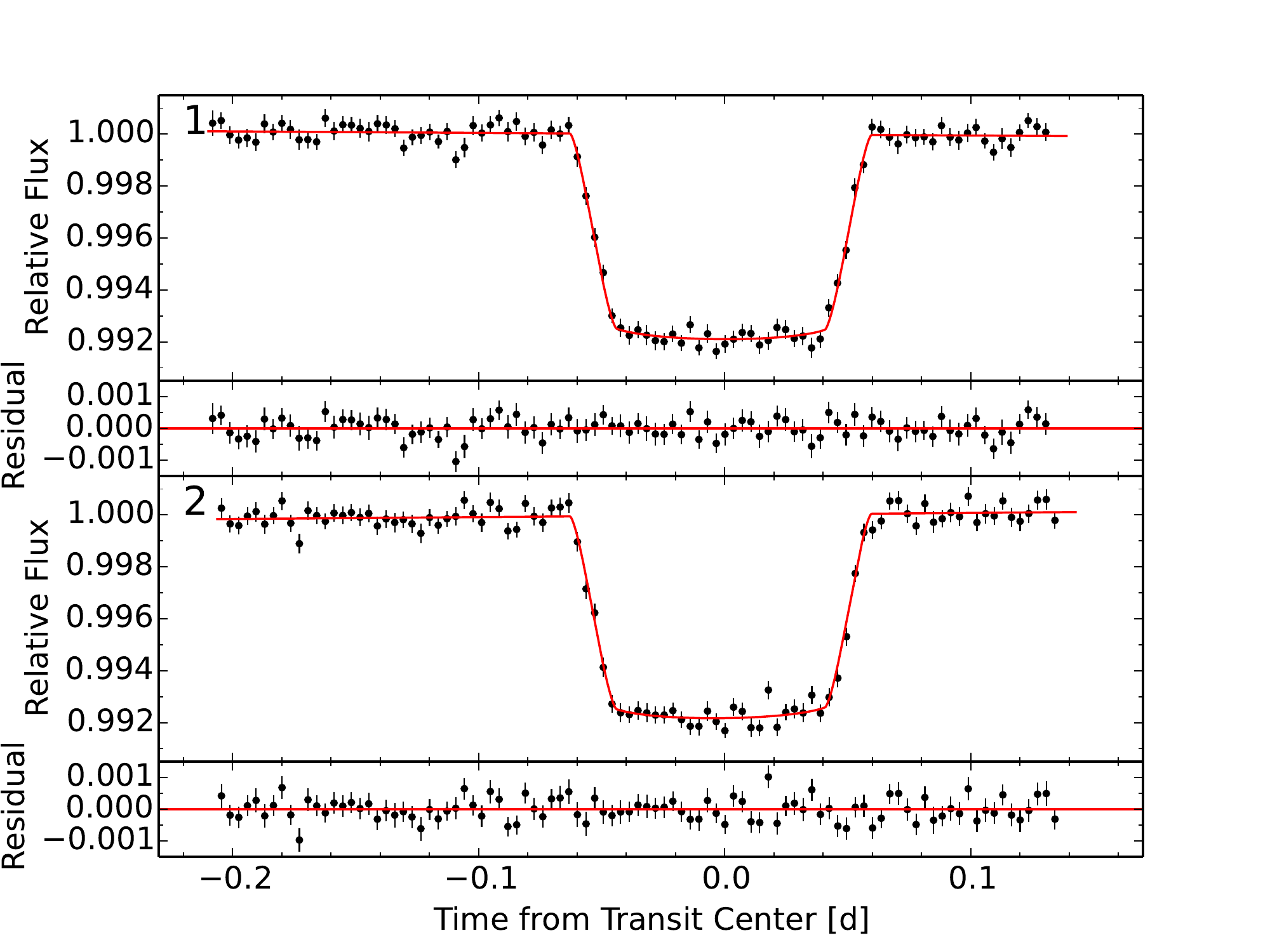}
\end{center}
\caption{Plot of the two 4.5~$\mu$m transit light curves, binned in three-minute intervals, with intrapixel sensitivity effects removed (black dots with error bars) and best-fit light curves overplotted (solid red lines). The residuals from each fit are below the corresponding light curve.} \label{transits}
\end{figure}

We fit the data for each of the twelve secondary eclipses and two transits, with intrapixel sensitivity correction calculated via pixel-mapping (Section~\ref{subsec:correction}), to our eclipse/transit model (Section~\ref{subsec:model}) using a Levenberg-Marquardt least-squares algorithm. The best-fit eclipse/transit parameters for the individual data sets are listed in Table~\ref{tab:values}; the center of eclipse/transit phase values were calculated using an updated orbital ephemeris derived from fitting all available transit times (Section~\ref{subsec:ephemeris}). We adjust the center of eclipse phases to correct for the 42.1~s delay (relative to the mid-transit time) due to the light travel time across the system \citep{loeb}. The individual secondary eclipse  and transit light curves with intrapixel sensitivity effects removed are shown in Figures~\ref{eclipses} and \ref{transits}, respectively, along with the corresponding residuals from the best-fit light curves. The standard deviations of the best-fit residuals for individual secondary eclipse and transit observations are higher than the predicted photon noise limit by a factor of 1.11 to 1.15.

We use two different methods to estimate the uncertainties in our best-fit parameters. First, we estimate the contribution of time-correlated noise to the uncertainty with the so-called ``prayer-bead'' (PB) method \citep{gillon}: After extracting the residuals from the best-fit solution, we group them into segments of length 14 and cyclically permute their order segment by segment 1000 times, each time adding the new residual series back to the best-fit solution and recomputing the parameters using the least-squares algorithm. For each of the three parameters, we create a histogram of all 1000 computed values and obtain the uncertainties from the 68$\%$ confidence limits. The choice of segment length here is set to be $\mathrm{floor}(n/1000)$, where $n$ is the number of data points (after outliers have been removed); we obtain consistent uncertainties when using different segment lengths. Second, we use a Markov chain Monte Carlo (MCMC) routine with $10^5$ steps to fit the parameters, setting the initial state to be the best-fit solution from the least-squares analysis and the uncertainty on individual points to be the standard deviation of the best-fit residuals. We discard an initial burn-in on each chain of length equal to 10$\%$ of the chain length, which we found ensured the removal of any initial transient behavior in a chain, regardless of the choice of initial state. The distribution of values for all parameters are roughly Gaussian and do not display any correlations with one another. As in the PB method, we set the uncertainties in the eclipse/transit parameters to be the 68$\%$ confidence limits. For both uncertainty calculation methods, the median of the resulting parameter value distribution lies within $\sim$0.1$\sigma$ of the best-fit value.

For each parameter, we choose the larger of the two errors and report it in Table~\ref{tab:values}. The MCMC method generally yields larger uncertainties for the parameters of individual eclipses than the PB method; the calculated MCMC errors in phase constant $c$ are larger for 9 out of 12 eclipses, while the MCMC errors in center of eclipse time $t_{0}$ and eclipse depth $d$ are larger for 8 out of 12 eclipses. The MCMC method yields larger uncertainties for the transit parameters in all cases except for the phase constant of transit 1.  Overall, the size of the PB errors for individual eclipse parameters range between 0.80 and 1.15 times that of the corresponding MCMC errors. In the case where the light curves have a significant component of red (i.e., time-correlated) noise, we would expect the PB uncertainties to be systematically larger than the corresponding MCMC errors \citep{carterwinn}. However, the characteristic time scale of variations in the residuals of individual eclipse/transit observations is typically on the order of an hour. As a result, cyclic permutation of residuals from an individual eclipse/transit observation, which has a length of around four hours, may not sufficiently sample the time-correlated noise signal, leading to random variations in the size of the PB errors of up to 20\% relative to the MCMC values.

\begin{table*}[t!]
\centering
\begin{threeparttable}
\caption{Global Best-fit Parameters} \label{tab:best}

\renewcommand{\arraystretch}{1.2}
\begin{center}
\begin{tabular}{c m{0.001cm} c m{0.001cm} c}
\hline\hline

Parameter\textsuperscript{a} &  &Value & & 68$\%$ Confidence Limits\\
\hline
Eclipse depth, $d^*$ ($\%$) & & 0.1580 & &+0.0033, $-$0.0039\\
Center of eclipse phase, $\phi^*$ & & 0.67004& & +0.00015, $-$0.00010\\
Phase constant, $c^*$ ($10^{-4}$~d$^{-1}$) & & 6.0& & +1.3, $-$1.6\\
Planet-to-star radius ratio, $R_{p}/R_{*}^*$ & & 0.08825 & &+0.00037, $-$0.00037 \\
Inclination, $i^*$ (deg) & & 84.11 & & +0.16, $-$0.16\\
Scaled semi-major axis, $a/R_{*}^*$ & & 7.052 & & +0.076, $-$0.097 \\

\hline\hline

\end{tabular}

\begin{tablenotes}
      \small
      \item \textsuperscript{a}The asterisks indicate that these parameter values are computed from the global fit of all twelve secondary eclipses and two transits.
    \end{tablenotes}

\end{center}
\end{threeparttable}
\end{table*}

\begin{figure}[t]
\begin{center}
\includegraphics[width=9.4cm]{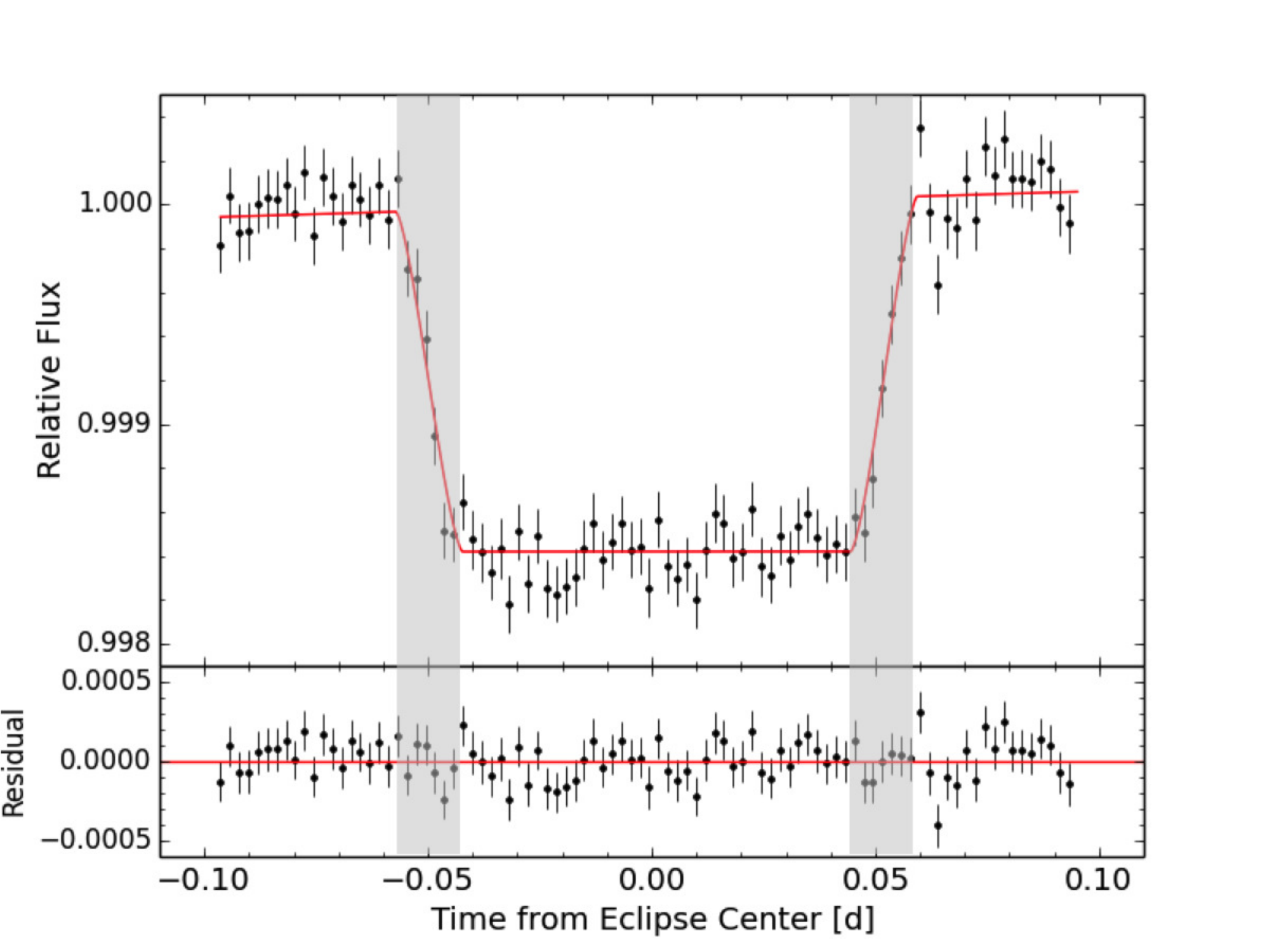}
\end{center}
\caption{Top panel: Combined secondary eclipse light curve from the global fit including data from all twelve eclipses, with intrapixel sensitivity effects removed, binned in three-minute intervals (black dots). The error bars show the standard deviation of all data points within each bin, divided by the square root of the number of points in each bin. The global best-fit light curve is overplotted (solid red line). The intervals of ingress and egress are highlighted in gray.  Bottom panel: Corresponding residuals from the best-fit solution.} \label{total}
\end{figure}

\begin{figure}
\begin{center}
\includegraphics[width=9.4cm]{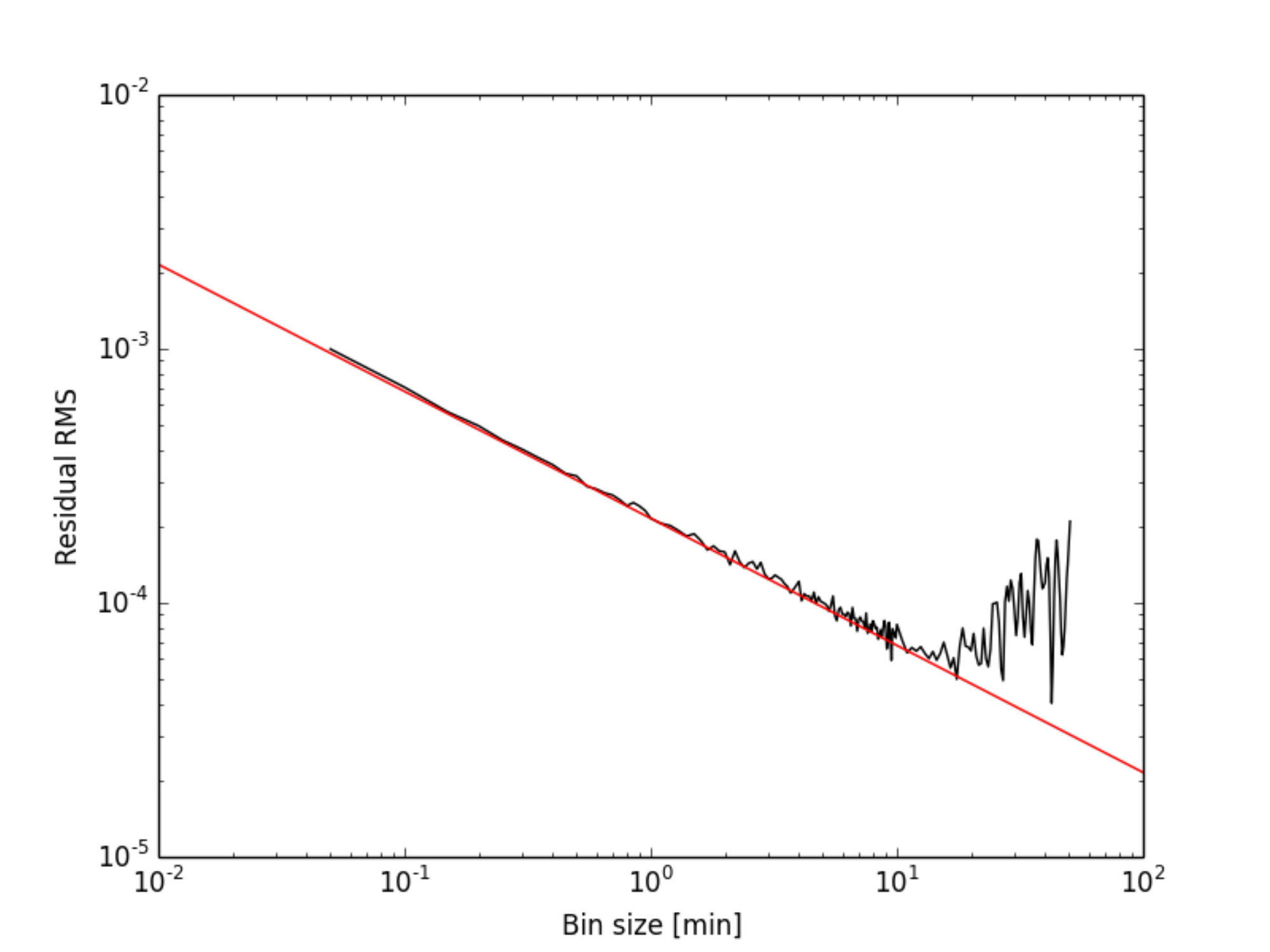}
\end{center}
\caption{Plot of the standard deviation of residuals from the best-fit solution to the combined light curve vs. bin size (black). The length of ingress and egress is 21.8 minutes. The solid red line denotes the inverse square-root dependence of white noise on bin size.} \label{rednoise}
\end{figure}

The global best-fit  parameters are determined by fitting all twelve secondary eclipses and two transits simultaneously. For each individual secondary eclipse data set, we convert the time series into a phase series and then define a global center of eclipse phase $\phi_{0}^*$. In the cumulative fit, the planet-to-star radius ratio $R_{p}/R_{*}^*$ is set as an additional global fit parameter that determines the depth of transit as well as the duration of ingress and egress. We also fit for global values of orbital inclination $i^*$ and scaled orbital semi-major axis $a/R_{*}^*$. The computed global eclipse parameter values are listed in Table~\ref{tab:best}.
The PB errors are up to 50\% larger than the MCMC errors with 200,000 steps for all the parameters except for the planet-to-star radius ratio and inclination, for which the PB errors are 30\% and 5\% smaller, respectively. As with the individual best-fit eclipse parameters, the larger of the two errors is reported for each parameter. The combined light curve with intrapixel sensitivity effects removed is shown in Figure~\ref{total}. We estimate the red noise by calculating the standard deviation of the residuals from the best-fit solution for various bin sizes, which is shown in Figure~\ref{rednoise} along with the inverse square-root dependence of white noise on bin size for comparison. We find that on timescales relevant for the eclipses (e.g., the length of ingress/egress  -- 21.8 minutes), the red noise contributes a relative scatter of roughly 0.01\%.

\section{Discussion}\label{sec:dis}
Fitting all twelve secondary eclipse and two transit observations simultaneously reduces the relative scatter in the binned residuals of the cumulative light curve (Figure~\ref{total}) to below the 0.05$\%$ level, which enables us to better discern any deviations from a spatially-uniform dayside brightness distribution at the time of secondary eclipse. At the same time, the large number of visits allows us to assess the orbit-to-orbit variability of the planet's atmosphere and to calculate improved estimates of the orbital parameters and ephemeris. We discuss each of these aspects separately below.

\subsection{Orbital parameters and ephemeris}\label{subsec:ephemeris}

In our global fit, we allowed the orbital inclination $i$ and scaled semi-major axis $a/R_{*}$ to vary as free parameters. These two quantities are important for secondary eclipse mapping because they are the primary determinants of the eclipse/transit duration as well as the length of ingress and egress, and small variations in their values can produce residuals during ingress and egress that mimic the expected signals from a planet with a dayside brightness distribution that is not spatially uniform. Conversely, setting the inclination and scaled semi-major axis as free parameters may cause our optimization routine to partially ``fit away'' any deviation in the light curve due to a genuine spatially-uniform dayside brightness. 

\begin{figure}[t!]
\begin{center}
\includegraphics[width=9.3cm]{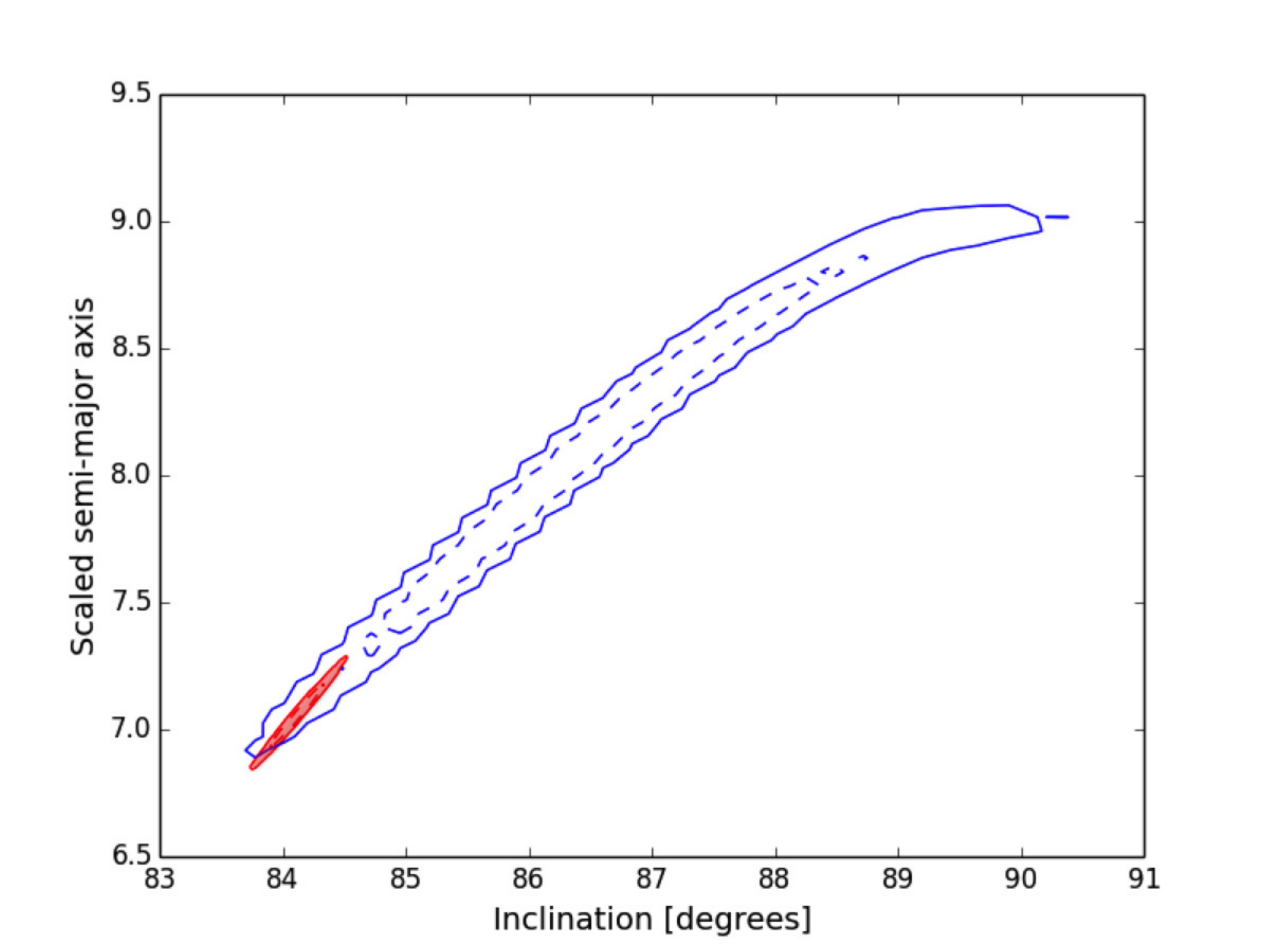}
\end{center}
\caption{Plot of the marginalized posterior probability distributions of inclination and scaled semi-major axis from the cumulative MCMC fit of the twelve secondary eclipses  with (red, filled contours) and without (blue, unfilled contours) the two transits included. The $1\sigma$ and $2\sigma$ contours are denoted by the dotted lines and solid lines, respectively.} \label{prob}
\end{figure}

We examine the effect of allowing inclination and scaled semi-major axis values to vary freely by utilizing a MCMC routine with 200,000 steps to calculate the best-fit values and $1\sigma$ uncertainties of the global eclipse parameters as well as $i$ and $a/R_{*}$, for cases where we fit (a) the twelve secondary eclipses alone and (b) all the secondary eclipses and transits together, as was done in the global fit in Section~\ref{subsec:fits}.  For the first case, we obtain $i = 86.47^{+1.67}_{-1.23}$~degrees and $a/R_{*} = 8.10^{+0.59}_{-0.54}$; for the second case, we obtain $i = 84.12\pm 0.16$~degrees and $a/R_{*} = 7.057\pm 0.090$. We see that these values are consistent at the $1.7\sigma$ level; the marginalized posterior probability distribution for inclination and scaled semi-major axis are shown in Figure~\ref{prob} with the $1\sigma$ and $2\sigma$ contours marked. 

It is evident that the inclusion of the transits in the cumulative fit places a much stronger constraint on the inclination and scaled semi-major axis, which underlines the advantage of including the two transits in our cumulative fit in Section~\ref{subsec:fits}. Meanwhile, the consistency between the distributions of inclination and scaled semi-major axis for the cases with and without the transits demonstrates that the twelve eclipses alone do appreciably constrain the value of these two parameters. Indeed, the values of the global eclipse depth and center of eclipse phase computed from the MCMC fits with and without transits included are consistent to well within $1\sigma$.

\begin{figure}[h]
\begin{center}
\includegraphics[width=9.3cm]{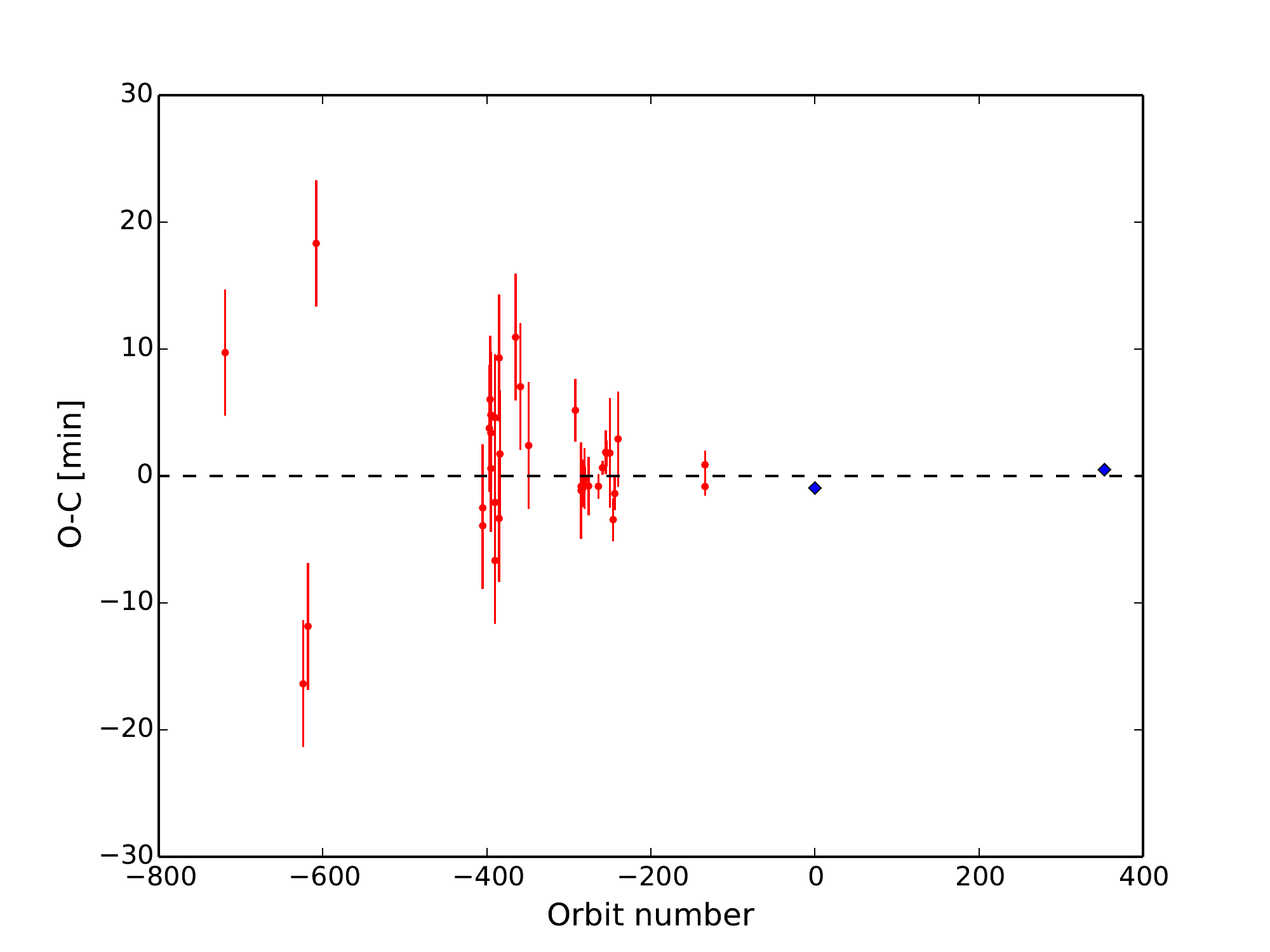}
\end{center}
\caption{Observed minus calculated transit times for all published observations (red circles -- previously-published) using the updated ephemeris calculated in Section~\ref{subsec:ephemeris}, including the two most recent transit times calculated in this paper (blue diamonds).} \label{octransits}
\end{figure}

We calculate an updated ephemeris for the XO-3 system using the transit times listed in Table~\ref{tab:values}, combined with all previously published values \citep[][]{johnskrull,hebrard,winn,winn2009,hirano}. We define the zeroth epoch as that of the first \textit{Spitzer} transit observation (transit 1 in this paper) and carry out a linear fit to all of the measured transit times. Our new observations extend the previous baseline by almost a factor of two, and as a result we derive a new, more precise estimate of the planet's orbital period and zeroth epoch midtransit time:
\begin{equation}\label{ephemeris}\begin{cases}P&=3.19153285 \pm 0.00000058~\mathrm{days}\\ T_{c,0}&=2455292.43266 \pm 0.00015~\mathrm{(BJD)}\end{cases} .\end{equation} 
The observed minus calculated transit times using these updated ephemeris values are plotted in Figure~\ref{octransits}

\begin{figure}[h]
\begin{center}
\includegraphics[width=9.3cm]{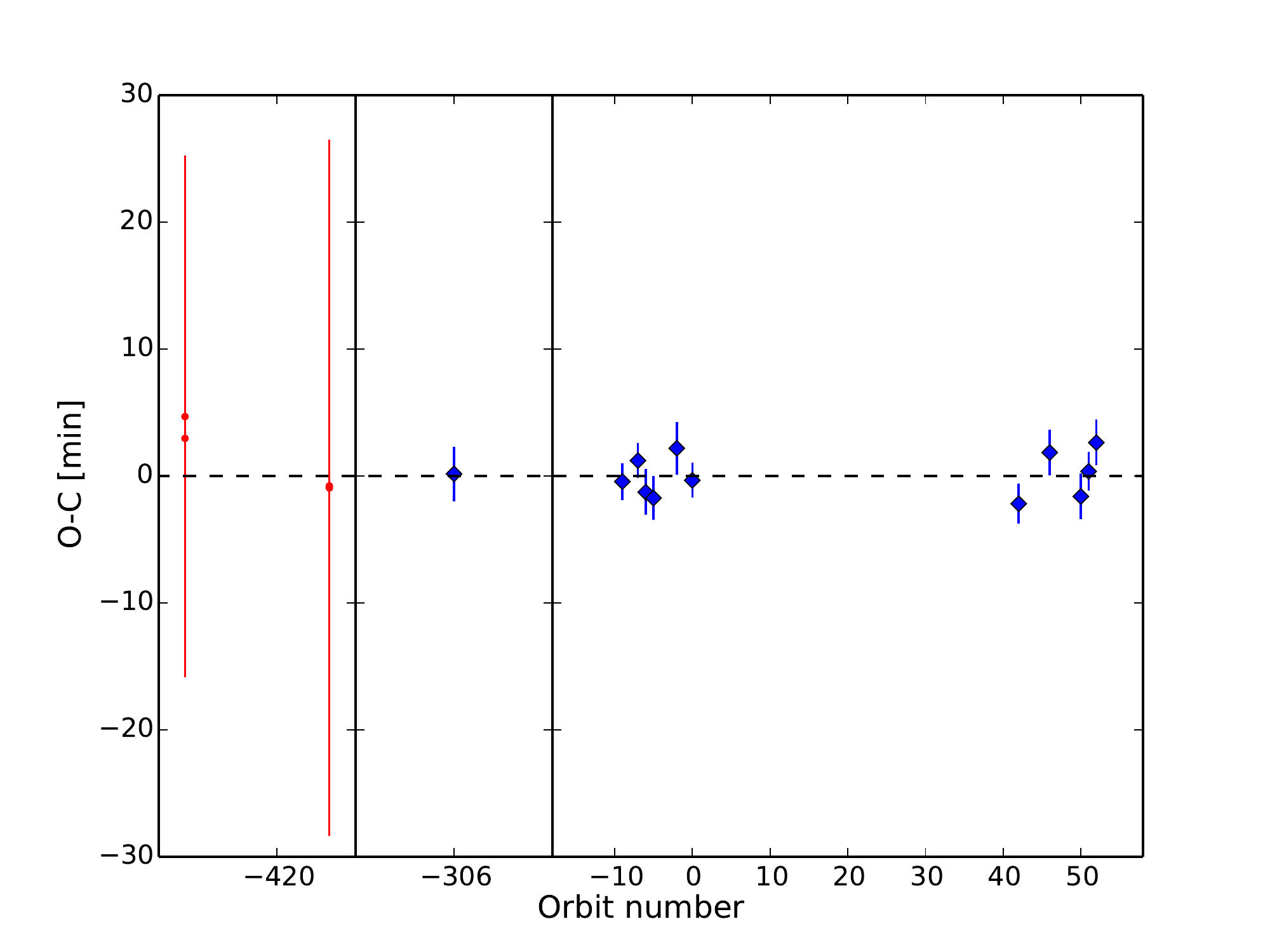}
\end{center}
\caption{Observed minus calculated secondary eclipse times for all published observations, with the twelve eclipse times computed in this paper denoted by blue diamonds and the eclipse times reported in \citet{machalek} denoted by red circles. The orbital ephemeris here is calculated using only the secondary eclipse times. The horizontal axis has been condensed for clarity.} \label{oceclipses}
\end{figure}

We obtain a second, independent estimate of the orbital period by fitting the secondary eclipse times calculated in \citet{machalek} as well as this paper and arrive at a best-fit value of $P=3.1915257\pm 0.0000041$~days. This period differs from the best-fit transit period by 2.1$\sigma$. The observed minus calculated secondary eclipse times are plotted in Figure~\ref{oceclipses}, where the zeroth epoch is defined here as the seventh eclipse analyzed in this work. 

We can derive limits on the periastron precession rate from the two different estimates of the orbital period. We use the formalism in \cite{pal} to convert the updated estimates for orbital eccentricity and pericenter longitude into a zeroth epoch center of eclipse time. Next, we introduce a $d\omega/dt$ term and calculate the predicted center of eclipse times at each of the other epochs of eclipse observations, which we then fit with a line to obtain a new value of the eclipse period. We limit the value of $d\omega/dt$ so that the resulting estimates of the eclipse period do not differ from the transit period calculated above by more than 3$\sigma$ and obtain the following constraint on the periastron precession rate:
\begin{equation}\label{precession}-8.5\times 10^{-4} < \frac{d\omega}{dt} < 2.9\times 10^{-3}~\mathrm{deg./day}.\end{equation}
For comparison, the expected periastron precession rate from general relativity and tides for the XO-3 system are $9.7\times 10^{-5}$ and $4.2\times10^{-5}$~deg./day, respectively \citep{jordan}. 

The most stringent limit we can place on the periastron precession rate is still roughly a factor of ten larger than the largest expected precession rate from theory. In order to reduce the derived limits to values comparable with the theoretical periastron precession rates, we would need to reduce the uncertainty in the eclipse period by roughly a factor of ten, which can be achieved by obtaining more secondary eclipse measurements of XO-3b over a sufficiently long time baseline. To assess this possibility, we envisioned a future campaign of eleven secondary eclipse observations with \textit{Spitzer}, spaced apart in a manner identical to that of the eleven most recent observations analyzed in this work (PID: 90032). Assuming that the timing of the future secondary eclipse observations match the predictions of our previously-calculated eclipse ephemeris and have uncertainties comparable with those in our data, we shifted the simulated future observations forward in time and calculated the new predicted uncertainty in orbital period. From this, we showed that a total baseline of about 15 years (versus the current baseline of roughly three years) is needed to detect the presence of periastron precession due to general relativity or tides.

\LongTables
\begin{deluxetable}{lrr}[t!]
    \tablecaption{XO-3b Radial velocity results}
    \tablehead{\colhead{Parameter} & \colhead{Value} & \colhead{Units}}
    \startdata
    
\sidehead{RV Model Parameters}$P_{b}$ & 3.19153247 $^{+5.5e-07}_{-5.4e-07}$ & days\\
$T_{c,b}$ & 2456419.04365 $\pm 0.00026$ & \bjdtdb\\
$e_{b}$ & 0.2769 $^{+0.0017}_{-0.0016}$ & \\
$\omega_{b}$ & 347.2 $^{+1.7}_{-1.6}$ & degrees\\
$K_{b}$ & 1478 $\pm 12$ & \ms\\
$\gamma_{1}$ & -203 $^{+18}_{-19}$ & \ms\\
$\gamma_{2}$ & 972 $\pm 70$ & \ms\\
$\gamma_{3}$ & -202 $^{+63}_{-60}$ & \ms\\
$\gamma_{4}$ & 185 $^{+19}_{-18}$ & \ms\\
$\gamma_{5}$ & -402 $\pm 48$ & \ms\\
$\dot{\gamma}$ & -0.023 $^{+0.026}_{-0.025}$ & \ms day$^{-1}$\\
jitter & 45.6 $^{+8.5}_{-7.0}$ & \ms\\
\sidehead{RV Derived Parameters}$e\cos{\omega}$ & 0.27005 $^{+0.00025}_{-0.00026}$ & \\
$e\sin{\omega}$ & -0.0612 $^{+0.0083}_{-0.0078}$ & \\

    \enddata\label{tab:rv}

    \end{deluxetable}

\begin{figure*}[t!]
\begin{center}
\includegraphics[width=18cm]{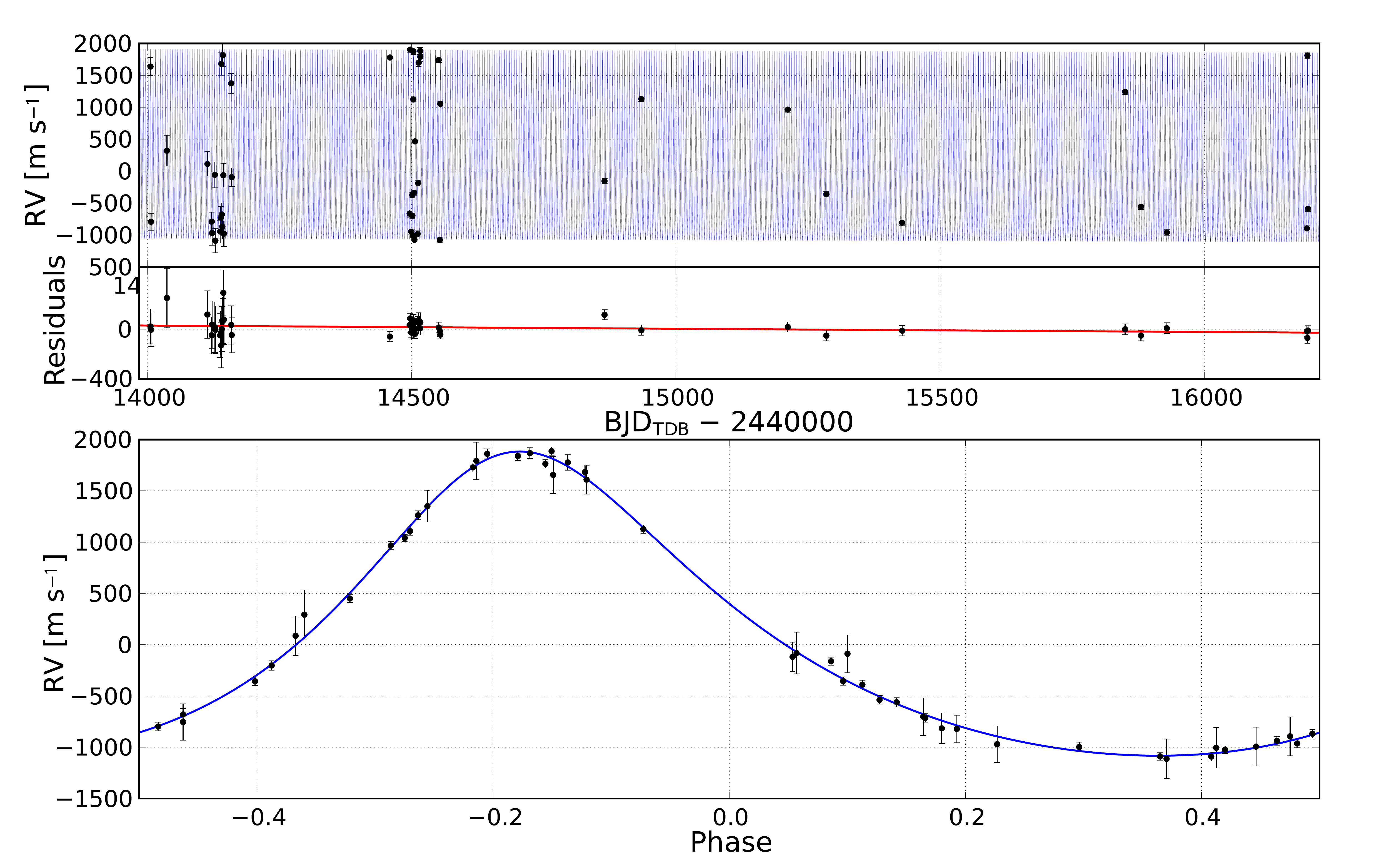}
\end{center}
\caption{Top panel: Full radial velocity fit of all published radial velocity measurements of XO-3. Middle panel: Corresponding residuals after the radial velocity solution for the transiting hot Jupiter is removed}. There is no significant linear acceleration detected in the data.  Bottom panel: Phased radial velocity curve. \label{fig:rv}
\end{figure*}

The measured center of eclipse times and updated transit ephemeris can be combined with the radial velocity measurements analyzed in \citet{knutson2014} to arrive at an updated estimate of the orbital eccentricity and pericenter longitude. Using the methodology of that paper, we obtain $e=0.2769^{+0.0017}_{-0.0016}$ and $\omega=347.2^{+1.7}_{-1.6}$~degrees. These values are consistent with the eccentricity and pericenter longitude estimates reported in \citet{knutson2014} ($e = 0.2833\pm0.0034$ and $\omega=346.8^{+1.6}_{-1.5}$~degrees). The updated values reduce the uncertainty in the eccentricity value by about a factor of two. The full results of our radial velocity fits are shown in Table~\ref{tab:rv} and Figure~\ref{fig:rv}. In addition to a new estimate of the orbital period ($P_{b}$) and center of transit time ($T_{c,b}$), we obtain values for the semi-amplitude of the planet's radial velocity ($K_{b}$), the relative radial velocity zero points for data collected by each of the different spectrographs from which radial velocity measurements of the system have been obtained ($\gamma_{1-5}$), the slope ($\dot{\gamma}$) of the best-fit radial velocity acceleration, as well as the radial velocity jitter. A complete description of the methodology used in our radial velocity fits can be found in \citet{knutson2014}.

\subsection{Atmospheric circulation}\label{subsec:circulation}

Examining the combined light curve and associated residual series in Figure~\ref{total}, we do not discern any anomalous signal in the residuals during ingress and egress above the level of the noise ($\sim$$0.05\%$). Therefore, we place an upper limit of 0.05\% on the relative magnitude of any deviation from a spatially-uniform surface brightness. We generate model secondary eclipse light curves using a one-dimensional semi-analytic model developed in \citet{cowanagol} and compare the observed upper limit on dayside brightness variations with the model-predicted signals in the residuals during ingress and egress. 

\begin{figure}[h!]
\begin{center}
\includegraphics[width=8.5cm]{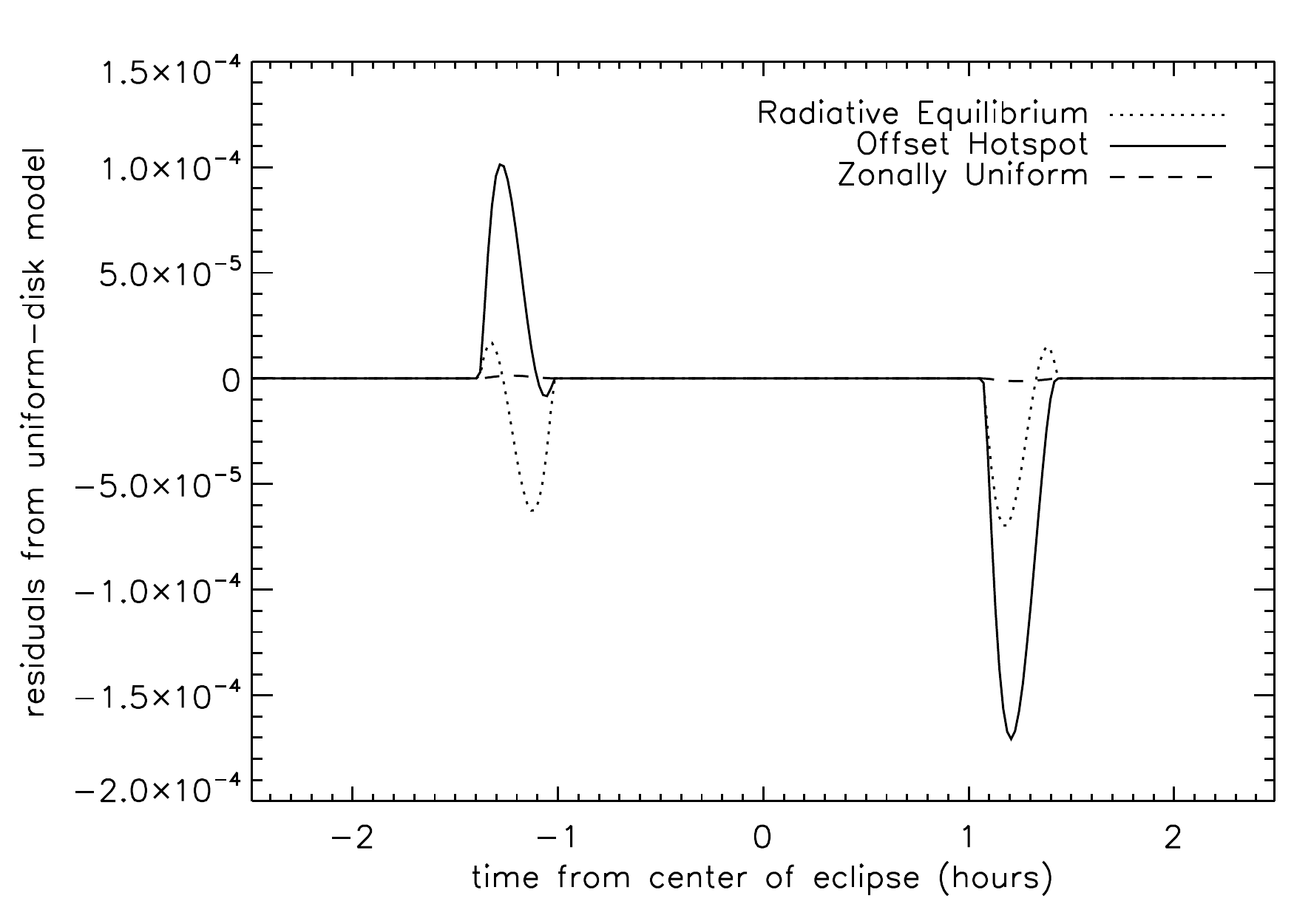}
\end{center}
\caption{Plot showing the predicted signals in the residuals for simulated light curves generated by the atmospheric model in \citet{cowanagol}. The case of radiative equilibrium corresponds to $\tau_{R}=0$ and $\omega_{rot}=0$, the case of offset hotspot corresponds to $\tau_{R}=0.1$~days and $\omega_{rot}=2.00$, and the case of zonally-uniform brightness corresponds to $\tau_{R}=100$~days and $\omega_{rot}=2.00$.} \label{models}
\end{figure}

\begin{figure}[t!]
\begin{center}
\includegraphics[width=9.3cm]{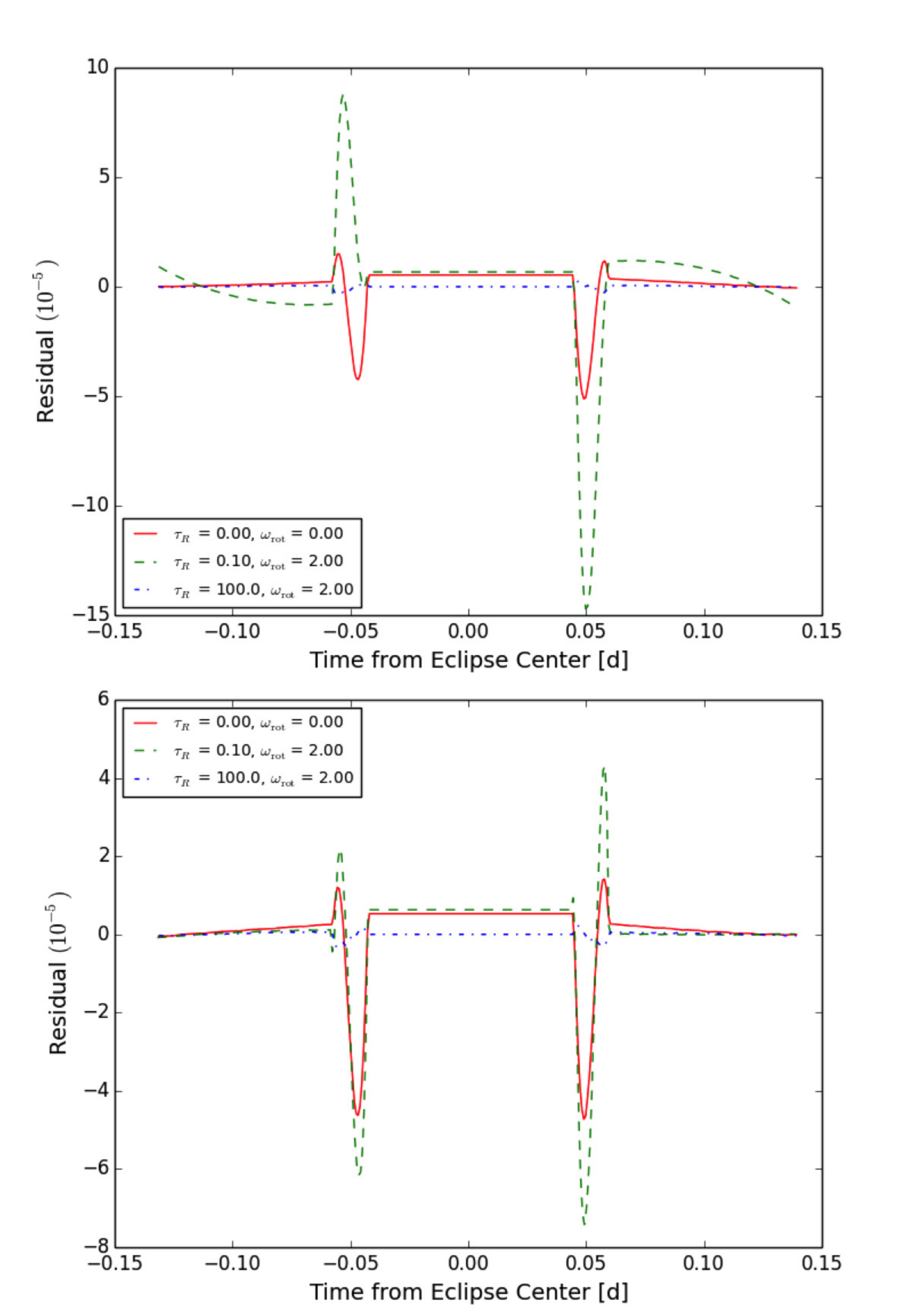}
\end{center}
\caption{Top: Plot of the residuals from fitting simulated light curves generated by the atmospheric model in \citet{cowanagol} with various input values of radiative timescale $\tau_{R}$ and atmospheric rotation speed $\omega_{rot}$ to our spatially-uniform disk brightness eclipse model, where the center of eclipse is fixed to the model value. Bottom: Same as above, but with the center of eclipse time as a free parameter.} \label{comparison}
\end{figure}

 The atmospheric model takes as inputs the radiative timescale $\tau_{R}$ and the rotational frequency of the planet's atmosphere $\omega_{rot}$ in units of the periastron orbital angular frequency. We  generate simulated light curves for models with (a) $\tau_{R}=0$ (radiative equilibrium; wind velocities are not important in this case and $\omega_{rot}$ is set to 0), (b) $\tau_{R}=0.1$~days and $\omega_{rot}=2.00$, which entails super-rotating winds and a hotspot that is offset from the substellar point, and (c) $\tau_{R}=100$~days and $\omega_{rot}=2.00$, which entails a zonally-uniform brightness. Figure~\ref{models} shows the predicted residuals  during ingress and egress from the simulated light curves. In order to determine the expected residuals during ingress and egress when fitting the simulated light curves with our uniform disk brightness eclipse model, we consider the case where we fix the center of eclipse time to the model value as well as the case where we let the center of eclipse time vary as a free parameter. In both cases, we allow the eclipse depth to vary and use a third-order polynomial to model the out-of-eclipse flux. The resulting residuals are shown in Figure~\ref{comparison}.

The magnitude of the residual signal during ingress and egress is reduced in both cases. In the case where the eclipse time is fixed, the least-squares algorithm adjusts the eclipse depth to partially compensate for the relatively large deviations from spatially-uniform brightness, resulting in a slight reduction in the magnitude of the expected residuals and non-zero residuals during eclipse. In the case where the eclipse time is allowed to vary, there is an additional reduction that most strongly affects the residuals from the model with large anti-symmetric residual signals during ingress and egress (i.e., $\tau_{R}=0.1$~days and $\omega_{rot}=2.00$). As discussed in \citet{dewit}, anti-symmetric residuals during ingress and egress can be ``fit away'' by adjusting the center of eclipse phase. In our cumulative fit of the twelve secondary eclipses, we have imprecise knowledge of the orbital parameters and must allow the eclipse phase to vary as a free parameter. As a result, the maximum expected relative magnitude of the residual signal during ingress and egress is $\sim$$8\times 10^{-5}$, much less than the noise level. Therefore, we cannot place meaningful constraints on the radiative timescale and wind speeds of XO-3b from these data.

\begin{figure}[t!]
\begin{center}
\includegraphics[width=9.3cm]{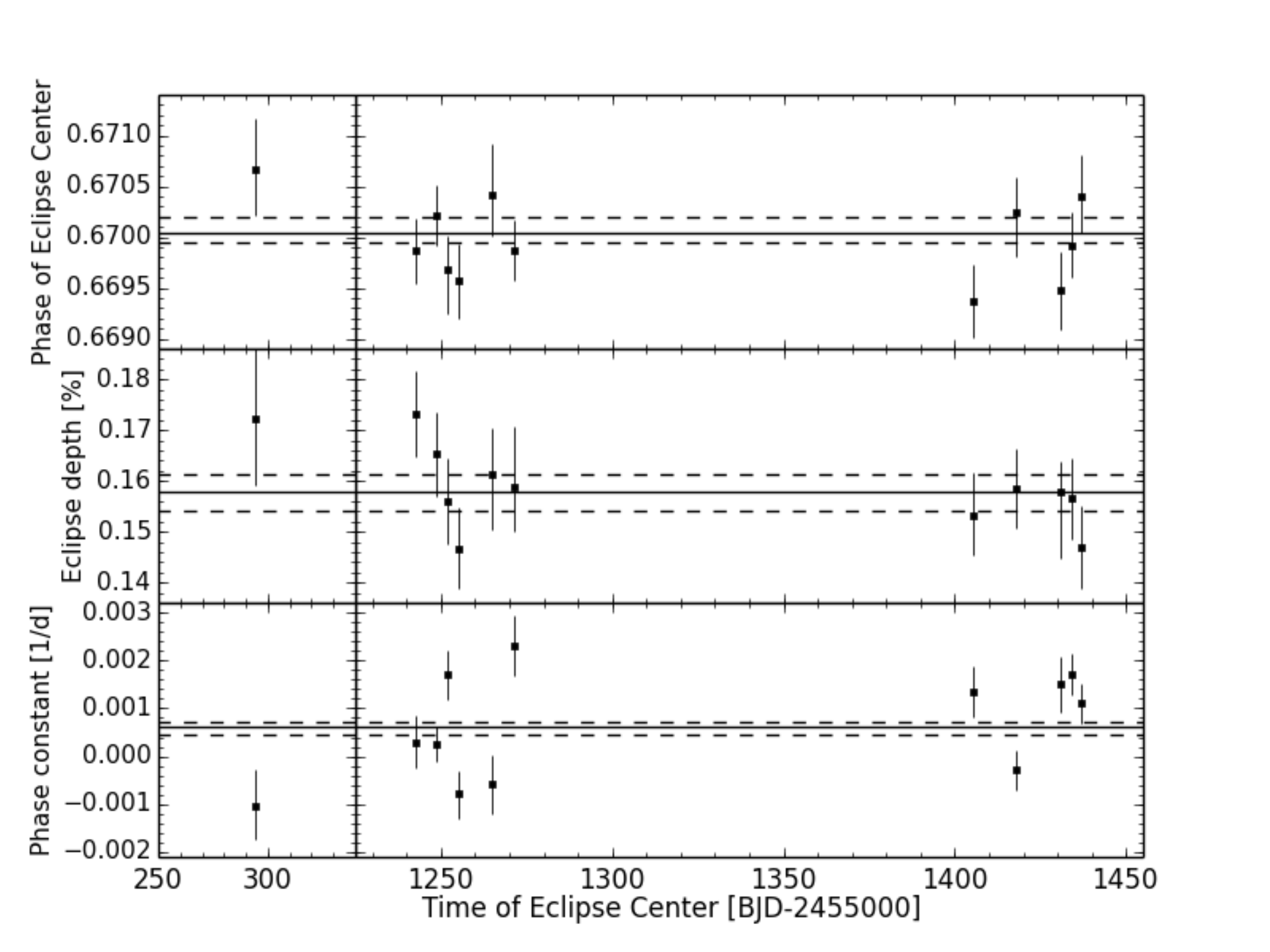}
\end{center}
\caption{Comparison plot of the best-fit center of eclipse phase, eclipse depth and phase constant for the twelve individual secondary eclipses (black dots) along with their uncertainties, and the global best-fit eclipse parameters (black lines) with their corresponding 68$\%$ confidence limits (dashed lines). The single early epoch measurement is from a partial phase curve observation on UT 2012 April 5-8 and was collected as part of a different observation program than the other eleven secondary eclipses.} \label{params}
\end{figure}

When fitting the secondary eclipse to the model light curve, we included the phase constant $c$ as a free parameter to describe variation in the observed out-of-eclipse flux due to the shape of the planet's phase curve in the region of the secondary eclipse. The global best-fit value, $c^*=6.0^{+1.3}_{-1.6} \times 10^{-4}~\mathrm{d}^{-1}$, is distinct from a flat phase curve slope at the $4\sigma$ level. It is generally difficult to estimate the slope of the phase curve from individual secondary eclipse observations, as this slope is degenerate with the change in flux due to the long-term position drift of the telescope. This is evident from the large scatter in phase constant values around the global best-fit value (see Figure~\ref{params}). However, by using all twelve secondary eclipse observations in the combined fit, we are able to break this degeneracy and obtain a unique estimate of the phase curve slope in the region of the secondary eclipse. 

The observed increase in the planet's brightness is likely due to an increase in the atmospheric temperature due to increasing stellar irradiation experienced by the planet during secondary eclipse. Another possible contributing factor may be changes in the apparent atmospheric brightness due to planetary rotation, which may be caused by an offset hotspot situated to the west of the substellar point during secondary eclipse. Such an offset hotspot can be generated by either westward winds or the presence of clouds east of the substellar point. Recent atmospheric circulation models do not readily produce westward winds that would lead to a westward offset of the hotspot at infrared wavelengths \citep{kataria}. While non-uniform clouds have been inferred from visible observations of Kepler-7b \citep{demory}, the significantly higher temperature and surface gravity of XO-3b place the equilibrium cloud decks at pressure levels greater than those probed by the 4.5-micron bandpass. Yet another possibility is non-uniform atmospheric chemical composition across the planet. In particular, variations in the abundance of CO$_{2}$ in the atmosphere of hot Jupiters has been shown to induce changes in the 4.5-micron flux from one side to the other, though the effect on the variation of emergent flux with phase is largely dampened by longitudinal homogenization of chemical abundances through circulation \citep{agundez}. Therefore, while all of these mechanisms may contribute to the brightening of XO-3b at the time of secondary eclipse, it is expected that the overall warming of the planet due to decreasing planet-star distance is the dominant factor. A full-orbit phase curve analysis of XO-3b is contained in a parallel study (Cowan et al.~2014, in prep.).

We can use the detection of brightening during eclipse as an independent constraint on the atmospheric properties. In particular, long radiative timescales tend to keep the planet from warming up as it approaches periastron. The increasing flux at superior conjunction, which occurs shortly before periastron, therefore indicates a relatively short radiative timescale. We generate simulated phase curves using the model in \citet{cowanagol} for a range of $\tau_{R}$ and $\omega_{rot}$ values. The positive global phase constant rules out radiative timescales longer than one day. This is consistent with the results of a phase curve analysis of another eccentric hot Jupiter, HAT-P-2b, which constrain the radiative timescale to $\tau_{R}<10$~hours \citep{lewis}.

\subsection{Orbit-to-orbit variability}

We next consider whether or not the planet displays any evidence for orbit-to-orbit variability in its dayside brightness. Recent work by \citet{parmentier} has suggested that measurable variations 
in the eclipse depth of a planet might result from orbit-to-orbit variations in the abundance of condensable species such 
as TiO or silicates. In Figure~\ref{params}, the best-fit eclipse parameters and uncertainties for the twelve individual eclipses are compared with the global best-fit eclipse parameters and uncertainties. The individual eclipse parameter values are largely consistent with the global values, and there is no apparent time-correlated variation in their values across the twelve eclipses.
We find an average deviation in the individual eclipse depths for XO-3b at 4.5~$\mu$m on the order of 5\%, 
which is roughly consistent with the measurement uncertainty. In \citet{machalek}, the eclipse depth at 4.5~$\mu$m was calculated to be $0.143\% \pm0.006\%$, which is consistent with our global eclipse depth at the 2.1$\sigma$ level. The agreement of individual eclipse depths measured at many different epochs suggests that any variability in the eclipse 
depth that is due to variability in XO-3b's global circulation patterns is likely small ($<5$\%).  
Recent three-dimensional (3D) atmospheric circulation models that only consider radiative, dynamical, and equilibrium chemical 
processes find fairly low levels ($<1$\%) of orbit-to-orbit variability in the atmospheres of circular \citep{showman} and 
eccentric \citep{lewis2010,kataria} exoplanets alike.  There are only two planets with comparable observational limits on their orbit-to-orbit variability: HD 189733b, a hot Jupiter on a circular orbit, and GJ 436b, which is a warm Neptune with a relatively modest orbital eccentricity \citep[$e\sim 0.15$;][]{knutson2014}. While the effects of condensation and/or turbulent mixing in the 
atmosphere of XO-3b may lead to some orbit-to-orbit variation in the eclipse depths, the relatively low level of variability evident in our data suggest that these processes do not appear to significantly alter the atmosphere's thermal structure on orbit-to-orbit (and longer) 
timescales  at the pressure levels probed by the 4.5~$\mu$m bandpass.

These same data can also be used to evaluate the reliability of the reported uncertainties in secondary eclipse depths measured by the \textit{Spitzer Space Telescope}. \citet{hansen} argue that errors in the eclipse parameters calculated from \textit{Spitzer} data may be significantly underestimated, a statement which has major implications for the ability to use secondary eclipse photometry to deduce characteristics of exoplanet atmospheres.  In particular, \citet{hansen} argue that individual secondary eclipse measurements do not fully account for detector systematics, which may lead to poor reproducibility of individual eclipse depth measurements and result in apparently inconsistent values for the eclipse depth from one study to the next. The 2.1$\sigma$ discrepancy between our global best-fit eclipse depth and the depth measured from the single 4.5~$\mu$m observation in \citet{machalek} is consistent with this purported tendency to underestimate the uncertainties of single-eclipse depths. However, the notable self-consistency of the individual depth measurements for the twelve secondary eclipses analyzed in our work  (Figure~\ref{params}) indicates that our technique of extracting photometry from the observations does adequately account for instrumental effects. Therefore, we expect that the reported uncertainties in the parameter values from both individual eclipse fits and the combined fit are an accurate representation of the measurement uncertainties. This is the first time that such an extensive data set has been collected in the short-wavelength (3.6 and 4.5 micron) \textit{Spitzer} channels, which are characterized by larger systematic effects due to the intrapixel sensitivity variations than the longer wavelength 5.8 and 8.0 micron channels.  Previous multiple-visit studies of HD 189733b and GJ 436b were both at 8.0 microns and were thus not sufficient to definitively assess the validity of the arguments in \citet{hansen}.

\section{Conclusion}
In this paper, we analyzed twelve secondary eclipse observations of the hot Jupiter XO-3b in the 4.5~$\mu$m \textit{Spitzer} band. After correcting for the intrapixel sensitivity effect, we fit each photometric time series with a uniform disk model light curve and measured the best-fit eclipse depths and center of eclipse times. We included two transits observed in the same band and fit all of the data simultaneously to obtain a global best-fit secondary eclipse depth of  $0.1580^{+0.0033}_{-0.0039}\%$ and a center of eclipse phase of $0.67004^{+0.00015}_{-0.00010}$, as well as updated values for orbital inclination ($i=84.11\pm 0.16$~degrees), planet-to-star radius ratio ($R_{p}/R_{*}=0.08825\pm 0.00037$), and scaled orbital semi-major axis ($a/R_{*}=7.052^{+0.076}_{-0.097}$). We combined the two transits analyzed in this work with all previously-published values and derived a more precise estimate of XO-3b's orbital period ($P=3.19153285\pm0.00000058$~days). By comparing the transit period with one derived from the secondary eclipse times, we were able to constrain the orbital periastron precession of the planet to between $-8.5\times 10^{-4}$ and $2.9\times 10^{-3}$~degrees per day. In addition, we incorporated the measured center of eclipse times and updated transit ephemeris in a radial velocity analysis to arrive at updated orbital eccentricity and pericenter longitude values of $e=0.2769^{+0.0017}_{-0.0016}$ and $\omega=347.2^{+1.7}_{-1.6}$~degrees, respectively.

The best-fit eclipse depths and center of eclipse times for individual secondary eclipse observations were found to be consistent with the corresponding global values. The lack of any discernible time-correlated variation in the eclipse depth values is indicative of a low level ($<5\%$) of orbit-to-orbit variability  in the planet's atmospheric brightness and is consistent with recent three-dimensional atmospheric circulation models \citep{showman,lewis2010}. Furthermore, the self-consistency of the individual best-fit eclipse parameter values derived from data collected over a span of more than three years demonstrates the reliability of multiple-epoch \textit{Spitzer} data and our ability to adequately account for instrumental effects. 

Our cumulative fit reduced the relative scatter in the binned residuals from the best-fit solution to less than 0.05\%, and we did not observe any signals in the residual series during ingress and egress above the noise level. We therefore conclude that any deviations from a spatially-uniform surface brightness on the planet's dayside must have a relative magnitude of less than 0.05\%. While the maximum expected residual signals during ingress and egress from plausible atmospheric models are smaller than the noise level achieved in our observations, our 4$\sigma$ detection of brightening at the time of eclipse enabled us to constrain the atmospheric radiative time scale of XO-3b to $\tau_{R}\lesssim 1$~day. More stringent constraints on XO-3b's orbital eccentricity from radial velocity fits could eventually yield predictions for the center of eclipse times that are precise enough to circumvent the problem encountered in our analysis, where the magnitude of the expected residual signal during ingress and egress was reduced by about an order of magnitude when we allowed the secondary eclipse time to vary in the fits. Meanwhile, an analysis of the available full-orbit phase curve of XO-3b could be combined with the results presented here to construct a more detailed picture of the planet's atmospheric properties. Lastly, a similarly-sized set of 4.5~$\mu$m \textit{Spitzer} secondary eclipse observations exists for another eccentric hot Jupiter, HAT-P-2b, which will provide a second, independent look at the issues of atmospheric circulation and variability in the eccentric orbit regime.


\small

\begin{thebibliography}{45}
\providecommand{\natexlab}[1]{#1}
\providecommand{\url}[1]{\texttt{#1}}
\expandafter\ifx\csname urlstyle\endcsname\relax
  \providecommand{\doi}[1]{doi: #1}\else
  \providecommand{\doi}{doi: \begingroup \urlstyle{rm}\Url}\fi

\bibitem[Agol et~al.(2010)Agol, Cowan, Knuston, Deming, Steffen, Henry, and
  Charbonneau]{agol}
E.~Agol, N.~B. Cowan, H.~A. Knuston, D.~Deming, J.~H. Steffen, G.~W. Henry, and
  D.~Charbonneau.
\newblock The climate of {HD 189733b} from fourteen transits and eclipses
  measured by {Spitzer}.
\newblock \emph{ApJ}, 721:\penalty0 1861, 2010.

\bibitem[Ag{\' u}ndez et~al.(2014)Ag{\' u}ndez, Parmentier, Venot, Hersant, and Selsis]{agundez}
M.~Ag{\' u}ndez, V.~Parmentier, O.~Venot, F.~Hersant, and F.~Selsis.
\newblock Pseudo {2D} chemical model of hot{-J}upiter atmospheres{:} application to {HD 209458b and HD 189733b}.
\newblock \emph{A\&A}, 564:\penalty0 A73, 2014.

\bibitem[Ballard et~al.(2010)Ballard, Charbonneau, Deming, Knuston,
  Christiansen, J., Fabrycky, Seager, and A'Hearn]{ballard}
S.~Ballard, D.~Charbonneau, D.~Deming, H.~A. Knuston, J.~L. Christiansen,
  Holman~M. J., D.~Fabrycky, S.~Seager, and M.~F. A'Hearn.
\newblock A search for a sub{-E}arth sized companion to {GJ 436} and a novel
  method to calibrate warm {S}pitzer {IRAC} observations.
\newblock \emph{PASP}, 122:\penalty0 1341, 2010.

\bibitem[Carter and Winn(2009)]{carterwinn}
J.~A. Carter and J.~N. Winn.
\newblock Parameter estimation from time{-}series data with correlated
  errors{:} {A} wavelet{-}based method and its application to transit light
  curves.
\newblock \emph{ApJ}, 704:\penalty0 51, 2009.

\bibitem[Charbonneau et~al.(2005)Charbonneau, Allen, Megeath, Torres, Alonso,
  Brown, Gilliland, Latham, Mandushev, O'Donovan, and
  Sozzetti]{charbonneau2005}
D.~Charbonneau, L.~E. Allen, S.~T. Megeath, G.~Torres, R.~Alonso, T.~M. Brown,
  R.~L. Gilliland, D.~W. Latham, G.~Mandushev, F.~T. O'Donovan, and
  A.~Sozzetti.
\newblock Detection of thermal emission from an extrasolar planet.
\newblock \emph{ApJ}, 626:\penalty0 523, 2005.

\bibitem[Charbonneau et~al.(2008)Charbonneau, Knutson, Barman, Allen, Mayor,
  Megeath, Queloz, and Udry]{charbonneau2008}
D.~Charbonneau, H.~A. Knutson, T.~Barman, L.~E. Allen, M.~Mayor, S.~T. Megeath,
  D.~Queloz, and S.~Udry.
\newblock Broadband emission spectrum of {HD} 189733b.
\newblock \emph{ApJ}, 686:\penalty0 1341, 2008.

\bibitem[Cowan and Agol(2011)]{cowanagol}
N.~B. Cowan and E.~Agol.
\newblock A model for thermal phase variations of circular and eccentric
  exoplanets.
\newblock \emph{ApJ}, 726:\penalty0 82, 2011.

\bibitem[de~Wit et~al.(2012)de~Wit, Gillon, Demory, and Seager]{dewit}
J.~de~Wit, M.~Gillon, B.-O. Demory, and S.~Seager.
\newblock Towards consistent mapping of distant worlds{:} {S}econdary{-}eclipse
  scanning of the exoplanet {HD 189733b}.
\newblock \emph{A\&A}, 548:\penalty0 A128, 2012.

\bibitem[Deming et~al.(2005)Deming, Seager, Richardson, and Harrington]{deming}
D.~Deming, S.~Seager, L.~J. Richardson, and J.~Harrington.
\newblock Infrared radiation from an extrasolar planet.
\newblock \emph{Nature}, 434:\penalty0 740, 2005.

\bibitem[Demory et~al.(2013)Demory, de Wit, Lewis, Fortney, Zsom, Seager, Knutson, Heng, Madhusudhan, Gillon, Barclay, Desert, Parmentier, and Cowan]{demory}
B.-O.~Demory, J.~de Wit, N.~Lewis, J.~Fortney, A.~Zsom, S.~Seager, H.~Knutson, K.~Heng, N.~Madhusudhan, M.~Gillon, T.~Barclay, J.-M.~Desert, V.~Parmentier, and N.~B. Cowan.
\newblock Inference of inhomogeneous clouds in an exoplanet atmosphere.
\newblock \emph{ApJL}, 776:\penalty0 L25, 2013.

\bibitem[Eastman et~al.(2010)Eastman, Siverd, and Gaudi]{eastman}
J.~Eastman, R.~Siverd, and B.~S. Gaudi.
\newblock Achieving better than 1 minute accuracy in the {H}eliocentric and
  {B}arycentric {J}ulian {D}ates.
\newblock \emph{PASP}, 122:\penalty0 935, 2010.

\bibitem[Fazio et~al.(2004)Fazio, Hora, Allen, Ashby, Barmby, Deutsch, Huang,
  Kleiner, Marengo, Megeath, Melnick, Pahre, Patten, Polizotti, Smith, Taylor,
  Wang, Willner, Hoffmann, Pipher, Forrest, McMurty, McCreight, McKelvey,
  McMurray, Koch, Moseley, Arendt, Mentzell, Marx, Losch, Mayman, Eichhorn,
  Krebs, Jhabvala, Gezari, Fixsen, Flores, Shakoorzadeh, Jungo, Hakun, Workman,
  Karpati, Kichak, Whitley, Mann, Tollestrup, Eisenhardt, Stern, Gorjian,
  Bhattacharya, Carey, Nelson, Glaccum, Lacy, Lowrance, Laine, Reach, Stauffer,
  Surace, Wilson, Wright, Hoffman, Domingo, and Cohen]{fazio}
G.~G. Fazio, J.~L. Hora, L.~E. Allen, M.~L.~N. Ashby, P.~Barmby, L.~K. Deutsch,
  J.-S. Huang, S.~Kleiner, M.~Marengo, S.~T. Megeath, G.~J. Melnick, M.~A.
  Pahre, B.~M. Patten, J.~Polizotti, H.~A. Smith, R.~S. Taylor, Z.~Wang, S.~P.
  Willner, W.~F. Hoffmann, J.~L. Pipher, W.~J. Forrest, C.~W. McMurty, C.~R.
  McCreight, M.~E. McKelvey, R.~E. McMurray, D.~G. Koch, S.~H. Moseley, R.~G.
  Arendt, J.~E. Mentzell, C.~T. Marx, P.~Losch, P.~Mayman, W.~Eichhorn,
  D.~Krebs, M.~Jhabvala, D.~Y. Gezari, D.~J. Fixsen, J.~Flores,
  K.~Shakoorzadeh, R.~Jungo, C.~Hakun, L.~Workman, G.~Karpati, R.~Kichak,
  R.~Whitley, S.~Mann, E.~V. Tollestrup, P.~Eisenhardt, D.~Stern, V.~Gorjian,
  B.~Bhattacharya, S.~Carey, B.~O. Nelson, W.~J. Glaccum, M.~Lacy, P.~J.
  Lowrance, S.~Laine, W.~T. Reach, J.~A> Stauffer, J.~A. Surace, G.~Wilson,
  E.~L. Wright, A.~Hoffman, G.~Domingo, and M.~Cohen.
\newblock The {Infrared Array Camera (IRAC)} for the \textit{Spitzer Space
  Telescope}.
\newblock \emph{ApJS}, 154:\penalty0 10, 2004.

\bibitem[Gillon et~al.(2009)Gillon, Smalley, Hebb, Anderson, Triaud, Hellier,
  Maxted, Queloz, and Wilson]{gillon}
M.~Gillon, B.~Smalley, L.~Hebb, D.~R. Anderson, A.~H. M.~J. Triaud, C.~Hellier,
  P.~F.~L. Maxted, D.~Queloz, and D.~M. Wilson.
\newblock Improved parameters for the transiting hot {J}upiters {WASP-4b} and
  {WASP-5b}.
\newblock \emph{A\&A}, 267:\penalty0 259, 2009.

\bibitem[Hansen et~al.(2014)Hansen, Schwartz, and Cowan]{hansen}
C.~J. Hansen, J.~C. Schwartz, and N.~B. Cowan.
\newblock Broadband eclipse spectra of exoplanets are featureless.
\newblock \emph{MNRAS}, accepted.

\bibitem[H{\' e}brard et~al.(2008)H{\' e}brard, Bouchy, Pont, Loeillet, Rabus,
  Bonfils, Moutou, Boisse, Delfosse, Desort, Eggenberger, Ehrenreich,
  Forveille, Lagrange, Lovis, Mayor, Pepe, Perrier, Queloz, Santos, S{\'
  e}gransan, Udry, and Vidal-Madjar]{hebrard}
G.~H{\' e}brard, F.~Bouchy, F.~Pont, B.~Loeillet, M.~Rabus, X.~Bonfils,
  C.~Moutou, I.~Boisse, X.~Delfosse, M.~Desort, A.~Eggenberger, D.~Ehrenreich,
  T.~Forveille, A.-M. Lagrange, C.~Lovis, M.~Mayor, F.~Pepe, C.~Perrier,
  D.~Queloz, N.~C. Santos, D.~S{\' e}gransan, S.~Udry, and A.~Vidal-Madjar.
\newblock Misaligned spin{-}orbit in the {XO-3} planetary system{?}
\newblock \emph{A\&A}, 488:\penalty0 763, 2008.

\bibitem[Heng and Showman(2014)]{heng}
K.~Heng and A.~P. Showman.
\newblock Atmospheric dynamics of exoplanets.
\newblock \emph{Annu. Rev. Earth Planet. Sci.}, submitted 2014.

\bibitem[Hirano et~al.(2011)Hirano, Narita, Sato, Winn, Aoki, Tamura, Taruya,
  and Suto]{hirano}
R.~Hirano, N.~Narita, B.~Sato, J.~N. Winn, W.~Aoki, M.~Tamura, A.~Taruya, and
  Y.~Suto.
\newblock Further observations of the tilted planet {XO-3}: {A} new
  determination of spin{–}orbit misalignment, and limits on differential
  rotation.
\newblock \emph{PASJ}, 63:\penalty0 L57, 2011.

\bibitem[Iro and Deming(2010)]{iro}
N.~Iro and L.~D. Deming.
\newblock A time{-}dependent radiative model for the atmosphere of the
  eccentric exoplanets.
\newblock \emph{ApJ}, 712:\penalty0 218, 2010.

\bibitem[Johns-Krull et~al.(2008)Johns-Krull, McCullough, Burke, Valenti,
  Janes, Heasley, Prato, Bissinger, Fleener, Foote, Garcia-Melendo, Gary,
  Howell, Mallia, Masi, and Vanmunster]{johnskrull}
C.~M. Johns-Krull, P.~R. McCullough, C.~J. Burke, J.~A. Valenti, K.~A. Janes,
  J.~N. Heasley, L.~Prato, R.~Bissinger, M.~Fleener, C.~N. Foote,
  E.~Garcia-Melendo, B.~L. Gary, P.~J. Howell, F.~Mallia, G.~Masi, and
  T.~Vanmunster.
\newblock {XO-3b:} {A} massive planet in an eccentric orbit transiting an {F5
  V} star.
\newblock \emph{ApJ}, 677:\penalty0 657, 2008.

\bibitem[Jord{\' a}n and Bakos(2008)]{jordan}
A.~Jord{\' a}n and G.~{\' A}. Bakos.
\newblock Observability of the general relativistic precession of periastra in
  exoplanets.
\newblock \emph{ApJ}, 685:\penalty0 543, 2008.

\bibitem[Kataria et~al.(2013)Kataria, Showman, Lewis, Fortney, Marley, and
  Freedman]{kataria}
T.~Kataria, A.~P. Showman, N.~K. Lewis, J.~J. Fortney, M.~S. Marley, and R.~S.
  Freedman.
\newblock Three{-}dimensional atmospheric circulation of hot {J}upiters on
  highly eccentric orbits.
\newblock \emph{ApJ}, 767:\penalty0 76, 2013.

\bibitem[Knutson et~al.(2008)Knutson, Charbonneau, Allen, Burrows, and
  Megeath]{knutson2008}
H.~A. Knutson, D.~Charbonneau, L.~E. Allen, A.~Burrows, and S.~T. Megeath.
\newblock The 3.6{-}8.0 {$\mu$}m broadband emission spectrum of {HD} 209458b{:}
  {E}vidence for an atmospheric temperature inversion.
\newblock \emph{ApJ}, 673:\penalty0 526, 2008.

\bibitem[Knutson et~al.(2010)Knutson, Howard, and Isaacson]{knutson2010}
H.~A. Knutson, A.~W. Howard, and H.~Isaacson.
\newblock A correlation between stellar activity and hot {J}upiter emission
  spectra.
\newblock \emph{ApJ}, 720:\penalty0 1569, 2010.

\bibitem[Knutson et~al.(2012)Knutson, Lewis, Fortney, Burrows, Showman, Cowan,
  Agol, Aigrain, Charbonneau, Deming, D{\' e}sert, Henry, Langton, and
  Laughlin]{knutson2012}
H.~A. Knutson, N.~K. Lewis, J.~J. Fortney, A.~Burrows, A.~P. Showman, N.~B.
  Cowan, E.~Agol, S.~Aigrain, D.~Charbonneau, D.~Deming, J.-M. D{\' e}sert,
  G.~W. Henry, J.~Langton, and G.~Laughlin.
\newblock 3.6 and 4.5 $\mu$m phase curves and evidence for non{-}equilibrium
  chemistry in the atmosphere of extrasolar planet {HD 189733b}.
\newblock \emph{ApJ}, 754:\penalty0 22, 2012.

\bibitem[Knutson et~al.(2014)Knutson, Fulton, Montet, Kao, Ngo, Howard, Crepp,
  Hinkley, Bakos, Batygin, Johnson, Morton, and Muirhead]{knutson2014}
H.~A. Knutson, B.~J. Fulton, B.~T. Montet, M.~Kao, H.~Ngo, A.~W. Howard, J.~R.
  Crepp, S.~Hinkley, G.~A. Bakos, K.~Batygin, J.~A. Johnson, T.~D. Morton, and
  P.~S. Muirhead.
\newblock Friends of {H}ot {J}upiters {I:} {A} radial velocity search of
  massive long{-}period companions to close{-}in gas giant planets.
\newblock \emph{ApJ}, 785:\penalty0 126, 2014.

\bibitem[Langton and Laughlin(2008)]{langton}
J.~Langton and G.~Laughlin.
\newblock Hydrodynamic simulations of unevenly irradiated {J}ovian planets.
\newblock \emph{ApJ}, 674:\penalty0 1106, 2008.

\bibitem[Lewis et~al.(2010)Lewis, Showman, Fortney, Marley, Freedman, and
  Lodders]{lewis2010}
N.~K. Lewis, A.~P. Showman, J.~J. Fortney, M.~S. Marley, R.~S. Freedman, and
  K.~Lodders.
\newblock Atmospheric circulation of eccentric hot {Neptune} {GJ436b}.
\newblock \emph{ApJ}, 720:\penalty0 344, 2010.

\bibitem[Lewis et~al.(2013)Lewis, Knuston, Showman, Cowan, Laughlin, Burrows,
  Deming, Crepp, Mighell, Agol, Bakos, Charbonneau, D{\' e}sert, Fischer,
  Fortney, Hartman, Hinkley, Howard, Johnson, Kao, Langton, and Marcy]{lewis}
N.~K. Lewis, H.~A. Knuston, A.~P. Showman, N.~B. Cowan, G.~Laughlin,
  A.~Burrows, D.~Deming, J.~R. Crepp, K.~J. Mighell, E.~Agol, G.~A. Bakos,
  D.~Charbonneau, J.-M. D{\' e}sert, D.~A. Fischer, J.~J. Fortney, J.~D.
  Hartman, S.~Hinkley, A.~W. Howard, J.~A. Johnson, M.~Kao, J.~Langton, and
  G.~W. Marcy.
\newblock Orbital phase variations of the eccentric giant planet {HAT-P-2b}.
\newblock \emph{ApJ}, 766:\penalty0 95, 2013.

\bibitem[Loeb(2005)]{loeb}
A.~Loeb.
\newblock A dynamical method of measuring the masses of stars with transiting
  planets.
\newblock \emph{ApJ}, 623:\penalty0 L45, 2005.

\bibitem[Machalek et~al.(2010)Machalek, Greene, McCullough, Burrows, Burke,
  Hora, Johns-Krull, and Deming]{machalek}
P.~Machalek, T.~Greene, P.~R. McCullough, A.~Burrows, C.~J. Burke, J.~L. Hora,
  C.~M. Johns-Krull, and D.~L. Deming.
\newblock Thermal emission and tidal heating of the heavy and eccentric planet
  {XO-3b}.
\newblock \emph{ApJ}, 711:\penalty0 111, 2010.

\bibitem[Madhusudhan et~al.(2014)Madhusudhan, Knutson, Fortney, and
  Barman]{madhusudhan}
N.~Madhusudhan, H.~Knutson, J.~J. Fortney, and T.~Barman.
\newblock Exoplanetary atmospheres.
\newblock In \emph{Protostars and Planets VI}, 2014.

\bibitem[Majeau et~al.(2012)Majeau, Agol, and Cowan]{majeau}
C.~Majeau, E.~Agol, and N.~B. Cowan.
\newblock A two-dimensional infrared map of the extrasolar planet {HD 189733b}.
\newblock \emph{ApJ}, 747:\penalty0 L20, 2012.

\bibitem[Mandel and Agol(2002)]{mandelagol}
K.~Mandel and E.~Agol.
\newblock Analytic lightcurves for planetary transit searches.
\newblock \emph{ApJ}, 580:\penalty0 L171, 2002.

\bibitem[Mighell(2005)]{mighell}
K.~J. Mighell.
\newblock Stellar photometry and astrometry with discrete point spread
  functions.
\newblock \emph{MNRAS}, 361:\penalty0 861, 2005.

\bibitem[O'Rourke et~al.(2014)O'Rourke, Knutson, Zhao, Fortney, Burrows, Agol,
  Deming, Howard, Lewis, Showman, and Todorov]{orourke}
J.~G. O'Rourke, H.~A. Knutson, M.~Zhao, J.~J. Fortney, A.~Burrows, E.~Agol,
  D.~Deming, J.-M. D{\' e}sert, A.~W. Howard, N.~K. Lewis, A.~P. Showman, and
  K.~O. Todorov.
\newblock Warm \textit{Spitzer} and {P}alomar near{-IR} secondary eclipse
  photometry of two {H}ot {J}upiters: {WASP-48b} and {HAT-P-23b}.
\newblock \emph{ApJ}, 781:\penalty0 109, 2014.

\bibitem[P{\' a}l et~al.(2010)P{\' a}l, Bakos, Torres, Noyes, Fischer, Johnson,
  Henry, Butler, Marcy, Howard, Sip{\" o}cz, Latham, and Esquerdo]{pal}
A.~P{\' a}l, G.~{\' A}. Bakos, G.~Torres, R.~W. Noyes, D.~A. Fischer, J.~A.
  Johnson, G.~W. Henry, R.~P. Butler, G.~W. Marcy, A.~W. Howard, B.~Sip{\"
  o}cz, D.~W. Latham, and G.~A. Esquerdo.
\newblock Refined stellar, orbital and planetary parameters for the eccentric
  {HAT-P-2} planetary system.
\newblock \emph{MNRAS}, 401:\penalty0 2665, 2010.

\bibitem[Parmentier et~al.(2013)Parmentier, Showman, and Lian]{parmentier}
V.~Parmentier, A.~P. Showman, and Y.~Lian.
\newblock {3D} mixing in hot {J}upiter atmospheres. {I.} {A}pplication to the
  day{/}night cold trap in {HD 209458b}.
\newblock \emph{A\&A}, 558:\penalty0 A91, 2013.

\bibitem[Perez-Becker and Showman(2013)]{perez-becker}
D.~Perez-Becker and A.~P. Showman.
\newblock Atmospheric heat redistribution on hot {J}upiters.
\newblock \emph{ApJ}, 776:\penalty0 134, 2013.

\bibitem[Petigura et~al.(2013)Petigura, Marcy, and Howard]{petigura}
E.~A. Petigura, G.~W. Marcy, and A.~W. Howard.
\newblock A plateau in the planet population below twice the size of {E}arth.
\newblock \emph{ApJ}, 770:\penalty0 69, 2013.

\bibitem[Rauscher et~al.(2007)Rauscher, Menou, Cho, Seager, and
  Hansen]{rauscher}
E.~Rauscher, K.~Menou, J.~Y.-K. Cho, S.~Seager, and B.~M.~S. Hansen.
\newblock Hot {J}upiter variability in eclipse depth.
\newblock \emph{ApJ}, 662:\penalty0 L115, 2007.

\bibitem[Showman et~al.(2009)Showman, Fortney, Lian, Marley, Freedman, Knuston,
  and Charbonneau]{showman}
A.~P. Showman, J.~J. Fortney, Y.~Lian, M.~S. Marley, R.~S. Freedman, H.~A.
  Knuston, and D.~Charbonneau.
\newblock Atmospheric circulation of hot {J}upiters: {C}oupled
  radiative{-}dynamical general circulation model simulations of {HD 189733b}
  and {HD 209458b}.
\newblock \emph{ApJ}, 699:\penalty0 564, 2009.

\bibitem[Sing(2010)]{sing}
D.~K. Sing.
\newblock Stellar limb{-}darkening coefficients for {CoRot} and {Kepler}.
\newblock \emph{A\&A}, 510:\penalty0 A21, 2010.

\bibitem[Todorov et~al.(2013)Todorov, Deming, Knutson, Burrows, Fortney, Lewis,
  Cowan, Agol, Desert, Sada, Charbonneau, Laughlin, Langton, and
  Showman]{todorov}
K.~O. Todorov, D.~Deming, H.~A. Knutson, A.~Burrows, J.~J. Fortney, N.~K.
  Lewis, N.~B. Cowan, E.~Agol, J.-M. Desert, P.~B. Sada, D.~Charbonneau,
  G.~Laughlin, J.~Langton, and A.~P. Showman.
\newblock \textit{Warm Spitzer} photometry of three {H}ot {J}upiters:
  {HAT-P-3b}, {HAT-P-4b} and {HAT-P-12b}.
\newblock \emph{ApJ}, 770:\penalty0 102, 2013.

\bibitem[Torres et~al.(2012)Torres, Fischer, Sozzetti, Buchhave, Winn, Holman,
  and Carter]{torres}
G.~Torres, D.~A. Fischer, A.~Sozzetti, L.~A. Buchhave, J.~N. Winn, M.~J.
  Holman, and J.~A. Carter.
\newblock Improved spectroscopic parameters for transiting planet hosts.
\newblock \emph{ApJ}, 757:\penalty0 161, 2012.

\bibitem[Visscher(2012)]{visscher}
C.~Visscher.
\newblock Chemical timescales in the atmospheres of highly eccentric
  exoplanets.
\newblock \emph{ApJ}, 757:\penalty0 5, 2012.

\bibitem[Williams et~al.(2006)Williams, Charbonneau, Cooper, Showman, and
  Fortney]{williams}
P.~K.~G. Williams, D.~Charbonneau, C.~S. Cooper, A.~P. Showman, and J.~J.
  Fortney.
\newblock Resolving the surfaces of extrasolar planets with secondary eclipse
  light curves.
\newblock \emph{ApJ}, 649:\penalty0 1020, 2006.

\bibitem[Winn et~al.(2008)Winn, Holman, Torres, McCullough, Johns-Krull,
  Latham, Shporer, Mazeh, Garcia-Melendo, Foote, Esquerdo, and Everett]{winn}
J.~N. Winn, M.~J. Holman, G.~Torres, P.~McCullough, C.~Johns-Krull, D.~W.
  Latham, A.~Shporer, T.~Mazeh, E.~Garcia-Melendo, C.~Foote, G.~Esquerdo, and
  M.~Everett.
\newblock The transit light curve project. {IX}. {E}vidence for a smaller
  radius of the exoplanet {XO-3b}.
\newblock \emph{ApJ}, 683:\penalty0 1076, 2008.

\bibitem[Winn et~al.(2009)Winn, Johnson, Fabrycky, Howard, Marcy, Narita,
  Crossfield, Suto, Turner, Esquerdo, and Holman]{winn2009}
J.~N. Winn, J.~A. Johnson, D.~Fabrycky, A.~W. Howard, G.~W. Marcy, N.~Narita,
  I.~J. Crossfield, Y.~Suto, E.~L. Turner, G.~Esquerdo, and M.~J. Holman.
\newblock On the spin{-}orbit misalignment of the {XO-3} exoplanetary system.
\newblock \emph{ApJ}, 700:\penalty0 302, 2009.

\end{thebibliography}
\end{document}